\documentclass[aps,prb,preprint,showpacs,superscriptaddress,amsmath]{revtex4-1}
\usepackage{txfonts}
\usepackage[dvips]{graphicx}
\usepackage{bm}
\usepackage{color}
\def\rnum#1{\expandafter{\romannumeral #1}} 
\def\Rnum#1{\uppercase\expandafter{\romannumeral #1}}

\newfont{\bg}{cmr10 scaled\magstep4}

\newcommand{\bigzerou}{\smash{\lower1.8ex\hbox{\bg 0}}}

\begin{document}
\title{Comprehensive Microscopic Theory for Coupling of Longitudinal--Transverse Fields and Individual--Collective Excitations}
\author{Tomohiro Yokoyama}
\email[E-mail me at: ]{tomohiro.yokoyama@mp.es.osaka-u.ac.jp}
\affiliation{Department of Material Engineering Science,
Graduate School of Engineering Science, Osaka University,
1-3 Machikaneyama, Toyonaka, Osaka 560-8531, Japan}
\author{Masayuki Iio}
\affiliation{Department of Material Engineering Science,
Graduate School of Engineering Science, Osaka University,
1-3 Machikaneyama, Toyonaka, Osaka 560-8531, Japan}
\affiliation{Department of Physics and Electronics,
Graduate School of Engineering, Osaka Prefecture University,
1-1 Gakuen-cho, Naka-ku, Sakai, Osaka 599-8531, Japan}
\author{Takashi Kinoshita}
\affiliation{Department of Physics and Electronics,
Graduate School of Engineering, Osaka Prefecture University,
1-1 Gakuen-cho, Naka-ku, Sakai, Osaka 599-8531, Japan}
\author{Takeshi Inaoka}
\affiliation{Department of Physics and Earth Sciences,
Faculty of Science, University of the Ryukyus,
1 Senbaru, Nishihara-cho, Okinawa 903-0213, Japan}
\author{Hajime Ishihara}
\affiliation{Department of Material Engineering Science,
Graduate School of Engineering Science, Osaka University,
1-3 Machikaneyama, Toyonaka, Osaka 560-8531, Japan}
\affiliation{Department of Physics and Electronics,
Graduate School of Engineering, Osaka Prefecture University,
1-1 Gakuen-cho, Naka-ku, Sakai, Osaka 599-8531, Japan}
\affiliation{Center for Quantum Information and Quantum Biology, Osaka University,
1-3 Machikaneyama, Toyonaka, Osaka 560-8531, Japan}

\date{\today}

\begin{abstract}
A plasmon is a collective excitation of electrons due to the Coulomb interaction.
Both plasmons and single-particle excitations (SPEs) are eigenstates of bulk metallic systems and they are orthogonal to each other.
In non-translationally symmetric systems such as nanostructures, plasmons, and SPEs coherently interact.
It has been well discussed that the plasmons and SPEs, respectively, can couple with
transverse (T) electric fields in such systems, and also that they are coupled with each other via longitudinal (L) field.
However, there has been a missing link in the previous studies:
the coherent coupling between the plasmons and SPEs mediated by the T field. 
Herein, we develop a theoretical framework to describe the self-consistent relationship between
the plasmons and SPEs through both the L and T fields.
The excitations are described in terms of the charge and current densities in a constitutive equation with
a nonlocal susceptibility, where the densities include the L and T components.
The electromagnetic fields originating from the densities are described in terms of the Green's function in the Maxwell's equations.
The T field is generated from both densities, whereas the L component is attributed to the charge density only.
We introduce a four-vector representation incorporating the vector and scalar potentials in the Coulomb gauge,
in which the T and L fields are separated explicitly.
The eigenvalues of the matrix for the self-consistent equations appear as the poles of the system excitations.
{
Numerical demonstration of the excitation spectrum is performed for a rectangular nanorod.
It indicates a non-negligible shift of the collective excitation and an enhancement of the energy transfer between
the excitations by the T-field-mediated interaction.
}
The developed formulation enables to approach unknown mechanisms for enhancement of the coherent coupling between
the plasmons and the hot carriers generated by radiative fields.
\end{abstract}
\maketitle

\section{INTRODUCTION}
\label{sec:intro}
Elucidation of the light--matter interaction is one of the core research topics of contemporary physics.
Recently, the behaviors of light-induced plasmons in metals have attracted considerable interest~\cite{Tame13}.
A plasmon is a quantum of collective electron motion due to the electron--electron interaction,
and has been clearly observed in experiment by Powell and Swan~\cite{Powell59}.
In bulk metals, a plasmon has a longitudinal (L) component only and, hence, cannot be excited by transverse (T) electromagnetic (EM) fields.
To induce a plasmon using light, the L component of the EM field, e.g., the evanescent mode on a surface, is required~\cite{book:NovotnyHecht}.
Cardinal methods to excite surface plasmon-polaritons (SPPs) on a 2D metal surface have been developed by
Otto~\cite{Otto68} and Kretschmann~\cite{Kretschmann68}.
The induced SPPs cause a charge--density spatial distribution on the surface, which in turn generates L fields.
Therefore, the EM fields and induced plasmons must be determined self-consistently~\cite{Feibelman82,Gerhardts84}.
In metal nanostructures, the SPPs are localized and enhance the L fields significantly.
Thus, SPPs are sensitive to surface states and are, therefore, applied in sensors for
gases, molecules, and biological matter~\cite{Garcia11,Zeng11,Gandhi19}.

One of the fascinating applications of light-induced plasmons is hot-carrier generation ~\cite{Govorov14,Brongersma15}
and injection to combined materials~\cite{White12,Govorov13,Kumar19},
which can be utilized for photocatalysis~\cite{Ueno16}, photodetection~\cite{Li17},
photocarrier injection~\cite{Tatsuma17}, and photovoltaics~\cite{Clavero14}.
The hot-carrier generation processes involving localized SPPs are categorized into two types:
those achieved through plasmon relaxation or through coherent coupling between
plasmons and individual single-particle excitations (SPEs).
Some theoretical studies have elucidated the relevant mechanisms of the plasmon relaxation approach. 
The first-principles calculation has been examined for hot-carrier generation on 2D metal surfaces~\cite{Sundararaman14,Bernardi15},
where the interband and intraband excitations make dominant and tunable contributions to the generation, respectively.
Hot-carrier generation for metallic nanostructures has also been investigated based on a phenomenological model with
several relaxation times~\cite{Zhang14,Ross15,Besteiro17} and using density functional theory~\cite{Manjavacas14}.
It was found that the nanostructure sizes and shapes influence the plasmon via
the electronic wavefunction and the enhanced electric field due to the hot spots. 
In those studies, a unidirectional energy transfer from the plasmon(-like) state to the hot carriers by
the electron SPEs was discussed~\cite{Zhang14,Ross15,Besteiro17,Manjavacas14,Chapkin18}.
As noted above, the second type of hot-carrier generation is due to coherent coupling between the plasmons and individual SPEs.
Recently, Ma {\it et al.} explicitly discussed the interplay between the plasmons and SPEs in
a nanocluster of Ag atoms using a real-time simulation based on time-dependent density functional theory~\cite{Ma15}.
The on- and off-resonant conditions between the plasmon and SPE were found to change the energy transfer processes.
For the off-resonant condition, a coherent Rabi oscillation between the plasmon and SPE was found.
You {\it et al.} also elucidated the bidirectional energy transfer between the plasmon and SPE
based on a model Hamiltonian with the Coulomb interaction~\cite{You18}, considering excited, injected, and extracted electrons.

In the present work, we focus on the fact that the coupling between the plasmons and SPEs via the T field,
as well as the treatment of the microscopic nonlocal response involving the T field, has been missed in the previous frameworks,
which is particularly related with the latter case in the above discussion. 
Recent ab initio calculations successfully implemented coherent coupling or hybridization between
plasmon-like and single-particle-like excitations~\cite{Zhang17,Chapkin18,Senanayake19,Fitzgerald17,Ma15,You18,Schira19}.
In those studies, the Coulomb interaction corresponding to part of the induced L field was considered.
However, the T field was not precisely considered despite the light-induced excitations.
As a phenomenological and semi-classical treatment of the microscopic nonlocal response involving the T field, 
the hydrodynamic model considering the ``pressure'' on the conduction electrons was applied to
the relation between the polarization (or current) and the electric
field~\cite{Pitarke07,McMahan09,Raza11,Toscano12,Moreau13,Wang13,Mortensen14,Christensen14}.
In such studies, aspects such as the plasmon peak shift, the presence of additional resonance above the plasma frequency,
and the size effect were well discussed~\cite{Raza11,Toscano12,Mortensen14,Christensen14}.
The importance of the nonlocal effect for a dimer nanostructure with a gap much narrower than
the light wavelength was noted~\cite{Toscano12,Mortensen14}.
However, depending on the sample structure or the situation, the hydrodynamic model requires
an additional boundary condition or additional treatments from outside the microscopic model~\cite{Svendsen20}.
Importantly, coherent coupling between the plasmons and SPEs has not yet been fully discussed with
this phenomenological model as a basis.
For a small nanostructure, the plasmon spectrum becomes indistinguishable from the SPE spectrum~\cite{Inaoka96}.
Therefore, for the nonlocal effect in nanostructures with radiative fields, a microscopic formulation must be developed.

Motivated by the above-mentioned situation, in this study, we develop a quantum mechanical framework for mesoscopic plasmonics,
focusing on the L and T components of EM fields and induced electronic responses in nanostructures that describe
light-induced plasmons (collective excitations), SPEs (individual excitations), and their coherent coupling via both the L and T fields.
An EM field induces charge polarization, charge current, spin fluctuation, and so on, and such electronic responses are
described by a constitutive equation with a nonlocal susceptibility.
The responses generate both L and T field components, and both field components must be comprehensively considered in
a self-consistent treatment of the Maxwell's and constitutive equations~\cite{book:cho1}.
Our formulation can be applied to arbitrary models of electronic states, such as those based on
the Drude model~\cite{Besteiro17}, jellium model with random phase approximation (RPA)~\cite{Prodan01},
first-principles calculation~\cite{Ma15}, or specific exotic materials, e.g., graphene~\cite{Wang13,Koppens11,Lu19}.
Such applications are expected to facilitate theoretical understanding of the interplay between
the individual and collective excitations in nanoscale systems.
{
Very recently, to understand the quantum effect in the light-matter interaction,
a combination of the density functional theory and the macroscopic Maxwell's equation has been discussed~\cite{Svendsen21}.
However, a discussion on the coupling of excitations caused by the L and T field components has not been developed yet.
In this study, we develop an understanding of the coherent coupling between the individual and collective excitations in terms of
the L and T fields and the charge and current densities.
To examine the feasibility of our formulation, we numerically demonstrate the excitation spectrum for a rectangular nanorod.
Then, we determine a non-negligible contribution of the T field component for a coherent coupling between
the individual and collective excitations and energy dissipation.
}
Further, our formulation provides a platform for the study of unrevealed physics concerning 
not only plasmons, but also excitons and/or other quasiparticles within nanostructures.

The remainder of this article is structured as follows.
In Section \ref{sec:Hamiltonian},
we describe the basic quantities used in our framework and the nanostructure Hamiltonian.
The Hamiltonian is separated into a nonperturbative term and an interaction with the field-induced potentials.
This separation determines the treatment of the optical response and the electronic states of the nanostructures.
In Section \ref{sec:susceptibility},
we introduce the four-vector representation and deduce the nonlocal susceptibility.
In Section \ref{sec:Maxwell}, we present a formal solution of the Maxwell's equations with vector and scalar potentials,
which is given by the Green's function describing the propagated fields from the excited current and charge densities.
In Section \ref{sec:SelfLT}, we explain the self-consistent treatment of the constitutive and Maxwell's equations demonstrated in
Secs.\ \ref{sec:susceptibility} and \ref{sec:Maxwell}.
{
In Section \ref{sec:RecRod}, we show the result of the numerical calculation of individual and collective excitations for
a rectangular nanorod to demonstrate practicality of our formulation.
}
In Section \ref{sec:discussion}, we discuss possible applications {and advantaged developments} of
our LT formulation for plasmonics and photonics.
Finally, we present the summary in Section \ref{sec:summary}.

\section{Basic quantities and Hamiltonian of electrons with electromagnetic field}
\label{sec:Hamiltonian}

In the present formulation, the fields are described by the vector and scalar potentials,
$\bm{A} (\bm{r},t)$ and $\phi (\bm{r},t)$, respectively.
In the quantum mechanics, they are essential rather than the electric and magnetic fields, $\bm{E}$ and $\bm{B}$.
It has been experimentally proven by the Aharonov--Bohm experiment~\cite{AB59,Tonomura86}.
To divide the fields into the L and T components, we apply the Coulomb gauge, ${\rm div} \bm{A} (\bm{r},t) = 0$,
where the vector potential describes the T field only.
In the constitutive equation, the electronic responses corresponding to the plasmon and SPE are described by
the charge and current densities, $\rho (\bm{r}, t)$ and $\bm{j} (\bm{r},t)$, respectively,
which are directly coupled with $\phi$ and $\bm{A}$ in the interaction Hamiltonian.
The nonlocal susceptibility is deduced from the Hamiltonian in accordance with
quantum linear response theory~\cite{Kubo57,book:cho1}, where the electron eigenstates determine the susceptibility.
Therefore, the nanostructure sizes, shapes, and internal structures strongly affect the optical response via the electron states.

To clearly define the light--matter interaction, we first discuss the matter and interaction Hamiltonians in terms of the potentials.
Within the nanostructures, mixing of the L and T field components and the densities occurs spontaneously.
To formulate the L and T components explicitly, we introduce a four-vector representation for the fields
{
$\bm{\mathcal{A}} = (\bm{A},-\phi /c)$
}
and densities $\bm{\mathcal{J}} = (\bm{j}, c\rho)$.
Hence, the nonlocal susceptibility is expressed as a $4 \times 4$ matrix.
This formulation enables us to exhibit the roles of the LT components of the fields and densities in
enhancing the optical response and the interplay between the plasmons and SPEs.

The Hamiltonian for the electrons in the EM fields is
\begin{equation}
\hat{H} = \sum_j 
\left[ \frac{\left\{ \hat{\bm{p}}_j - e \int d\bm{r} \bm{A} (\bm{r},t)
\delta (\bm{r} - \hat{\bm{r}}_j) \right\}^2}{2m^*} - \varepsilon_{\rm F} \right]
+ e \sum_j  \int d\bm{r} \phi (\bm{r},t) \delta (\bm{r} - \hat{\bm{r}}_j)
+ \frac{1}{2} \sum_{i \ne j} \frac{e^2}{4\pi\varepsilon_0 |\hat{\bm{r}}_i - \hat{\bm{r}}_j|},
\label{eq:Hamiltonian}
\end{equation}
where $e$ is the charge of electron and $\varepsilon_0$ is the dielectric constant in vacuum.
$\varepsilon_{\rm F}$ is the Fermi energy and $m$ is the (effective) mass for the conduction electrons.
The $\hat{ \ }$ symbol indicates the electron operator.
In the case of the Coulomb gauge, ${\rm div} \bm{A} = 0$, and
$\phi (\bm{r},t) = \phi_{\rm ncl} (\bm{r},t) + \phi_{\rm ext} (\bm{r},t)$ in the second term
describes the L field due to the nuclei and external origins.
The third term is the electron--electron (Coulomb) interaction, which is also longitudinal.

The first term is expanded as
\begin{eqnarray}
\frac{\left\{ \hat{\bm{p}}_j - e \int d\bm{r} \bm{A} (\bm{r},t) \delta (\bm{r} - \hat{\bm{r}}_j) \right\}^2}{2m^*}
&=& \frac{\hat{\bm{p}}_j^2}{2m^*}
- \frac{e}{2m^*} \int d\bm{r}
\left( \hat{\bm{p}}_j \delta (\bm{r} - \hat{\bm{r}}_j) + \delta (\bm{r} - \hat{\bm{r}}_j) \hat{\bm{p}}_j \right)
\cdot \bm{A} (\bm{r},t) \nonumber \\
& & \hspace{30mm}
+ \frac{e^2}{2m^*} \int d\bm{r} \delta (\bm{r} - \hat{\bm{r}}_j) \bm{A} (\hat{\bm{r}}_j,t) \cdot \bm{A} (\bm{r},t).
\label{eq:peA}
\end{eqnarray}
With the introduction of a current operator
\begin{equation}
\hat{\bm{I}} (\bm{r}) \equiv \frac{e}{2m^*} \sum_j
\left( \hat{\bm{p}}_j \delta (\bm{r} - \hat{\bm{r}}_j) + \delta (\bm{r} - \hat{\bm{r}}_j) \hat{\bm{p}}_j \right),
\label{eq:Iop}
\end{equation}
the second term on the right-hand side (r.h.s.) of Eq.\ (\ref{eq:peA}) gives the light--matter interaction
\begin{equation}
\hat{H}_{IA} = - \int d\bm{r} \hat{\bm{I}} (\bm{r}) \cdot \bm{A} (\bm{r},t).
\label{eq:H_int}
\end{equation}
Then, the Hamiltonian can be separated as follows:
\begin{equation}
\hat{H} = \hat{H}_0 + \left( \hat{H}_{IA} + \hat{H}_{A^2} + \hat{H}_{\rm ext} \right),
\label{eq:H0Hml}
\end{equation}
with
\begin{equation}
\hat{H}_0 = \sum_j \left[ \frac{\hat{ \bm{p}}_j^2}{2m^*} - \varepsilon_{\rm F} \right]
+ e \sum_j \int d\bm{r} \phi_{\rm ncl} (\bm{r},t) \delta (\bm{r} - \hat{\bm{r}}_j)
+ \frac{1}{2} \frac{e^2}{4\pi\varepsilon_0} \sum_{i \ne j} \frac{1}{|\hat{\bm{r}}_i - \hat{\bm{r}}_j|}.
\label{eq:Hmatter}
\end{equation}
Note that the summations over $i,j$ include the electron spin degrees of freedom.

The second quantization of these Hamiltonians is
\begin{eqnarray}
\hat{H}_0
&=& \int d\bm{x} \hat{\Psi}^\dagger (\bm{x})
\left( -\frac{\hbar^2 \bm{\nabla}_x^2}{2m^*} - \varepsilon_{\rm F} \right) \hat{\Psi} (\bm{x})
+ e \int d\bm{x} \hat{\Psi}^\dagger (\bm{x})
 \int d\bm{r} \phi_{\rm ncl} (\bm{r},t) \delta (\bm{r} - \bm{x}) \hat{\Psi} (\bm{x}) \nonumber \\
& & \hspace{20mm} + \frac{1}{2} \frac{e^2}{4\pi\varepsilon_0}
\int d\bm{x} \int d\bm{x}^\prime \hat{\Psi}^\dagger (\bm{x}) \hat{\Psi}^\dagger (\bm{x}^\prime)
\frac{1}{|\bm{x} - \bm{x}^\prime|} \hat{\Psi} (\bm{x}^\prime) \hat{\Psi} (\bm{x})
\nonumber \\
&=& \sum_{n,n^\prime} \int d\bm{x} \psi_{n^\prime}^* (\bm{x})
\left( -\frac{\hbar^2 \bm{\nabla}_x^2}{2m^*} - \varepsilon_{\rm F} \right) \psi_n (\bm{x}) \hat{a}_{n^\prime}^\dagger \hat{a}_n
+ \sum_{n,n^\prime} \int d\bm{x} \rho_{n^\prime n} (\bm{x}) \phi_{\rm ncl} (\bm{x},t)
\hat{a}_{n^\prime}^\dagger \hat{a}_n \nonumber \\
& & \hspace{40mm}
+ \frac{1}{2} \sum_{n,n^\prime,m,m^\prime} \int d\bm{x} \int d\bm{x}^\prime 
\frac{\rho_{n^\prime n} (\bm{x}) \rho_{m^\prime m} (\bm{x}^\prime)}{4\pi\varepsilon_0 |\bm{x} - \bm{x}^\prime|}
\hat{a}_{n^\prime}^\dagger \hat{a}_{m^\prime}^\dagger \hat{a}_m \hat{a}_n,
\label{eq:2ndH_0} \\
\hat{H}_{IA}
&=& - \sum_{n,n^\prime} \int d\bm{x} \bm{I}_{n^\prime n} (\bm{x}) \cdot \bm{A} (\bm{x},t)
\hat{a}_{n^\prime}^\dagger \hat{a}_n,
\label{eq:2ndH_IA} \\
\hat{H}_{A^2}
&=& \frac{e^2}{2m^*} \sum_{n,n^\prime} \int d\bm{x} \psi_{n^\prime}^* (\bm{x})
\bm{A} (\bm{x},t) \cdot \bm{A} (\bm{x},t) \psi_n (\bm{x}) \hat{a}_{n^\prime}^\dagger \hat{a}_n,
\label{eq:2ndH_A2} \\
{
\hat{H}_{\rm ext}
}
&=&
{
\sum_{n,n^\prime} \int d\bm{x} \rho_{n^\prime n} (\bm{x}) \phi_{\rm ext} (\bm{x},t)
\hat{a}_{n^\prime}^\dagger \hat{a}_n,
}
\label{eq:2ndH_ext}
\end{eqnarray}
where $\hat{\Psi} (\bm{x}) = \sum_n \psi_n (\bm{x}) \hat{a}_n$ is the field operator.
Note that the $n$ state includes the spin degrees of freedom.
Here,
\begin{equation}
\rho_{n^\prime n} (\bm{x}) = e \psi_{n^\prime}^* (\bm{x}) \psi_n (\bm{x})
\label{eq:charge}
\end{equation}
indicates an element of charge density and
\begin{equation}
\bm{I}_{n^\prime n} (\bm{x}) = - \frac{e\hbar}{2im^*}
\Big[ \left\{ \bm{\nabla}_x \psi_{n^\prime}^* (\bm{x}) \right\} \psi_n (\bm{x})
 - \psi_{n^\prime}^* (\bm{x}) \left\{ \bm{\nabla}_x \psi_n (\bm{x}) \right\} \Big]
\label{eq:currentI}
\end{equation}
is derived from the current operator in Eq.\ (\ref{eq:Iop}).
Note that $\rho_{n^\prime n} (\bm{x})$ has to follow the continuous relation with the charge current density.
Further, the single-electron wave function follows the time-dependent Schr\"odinger equation,
\begin{equation}
i\hbar \frac{\partial \psi (\bm{x},t)}{\partial t}
= \left[ -\frac{\hbar^2}{2m^*} \left( \bm{\nabla}_x - \frac{ie}{\hbar} \bm{A} (\bm{x},t) \right)^2
+ e \phi_{\rm ncl} (\bm{x},t) \right] \psi (\bm{x},t).
\end{equation}
Then, $\rho_{n^\prime n} (\bm{x})$ satisfies
\begin{equation}
\frac{\partial \rho_{n^\prime n} (\bm{x},t)}{\partial t}
= - \bm{\nabla}_x \cdot \bm{j}_{n^\prime n} (\bm{x},t),
\label{eq:continuous}
\end{equation}
where
\begin{equation}
\bm{j}_{n^\prime n} (\bm{x},t) = \bm{I}_{n^\prime n} (\bm{x})
- \frac{e^2}{m^*} \psi_{n^\prime}^* (\bm{x}) \bm{A} (\bm{x},t) \psi_n (\bm{x})
\label{eq:currentJ}
\end{equation}
is the modified current density in the EM field.
The second term contributes to the $O(\bm{A}^2)$ term.
The light--matter interaction component with the vector potential in Eq.\ (\ref{eq:H0Hml}) is approximately
\begin{eqnarray}
\hat{H}_{IA} + \hat{H}_{A^2}
&\simeq& - \sum_{n,n^\prime} \int d\bm{x} \bm{j}_{n^\prime n} (\bm{x},t) \cdot \bm{A} (\bm{x},t)
\hat{a}_{n^\prime}^\dagger \hat{a}_n \nonumber \\
&=& - \int d\bm{x} \hat{\bm{j}} (\bm{x},t) \cdot \bm{A} (\bm{x},t)
\equiv \hat{H}_{\rm int}.
\label{eq:HJA}
\end{eqnarray}

The second and third terms on the r.h.s. of Eq.\ (\ref{eq:2ndH_0}) can be interpreted as
the L component of the electric fields due to internal sources.
In a following formulation of the constitutive equation, the T and L fields are explicitly distinguished by
the vector and scalar potentials, respectively, with the Coulomb gauge.
To describe the susceptibility for the vector and scalar potentials,
we treat the many-electron system with $\hat{H}_0$ on a one-electron basis
and rewrite the second and third terms as
\begin{eqnarray}
& & \int d\bm{x} \sum_{n,n^\prime} \hat{a}_{n^\prime}^\dagger
\rho_{n^\prime n} (\bm{x},t) \left[
\phi_{\rm ncl} (\bm{x},t)
+ \frac{e}{4\pi\varepsilon_0} \int d\bm{x}^\prime
\frac{\hat{\Psi}^\dagger (\bm{x}^\prime,t) \hat{\Psi} (\bm{x}^\prime,t)}{|\bm{x} - \bm{x}^\prime|}
\right] \hat{a}_n \nonumber \\
& & = \int d\bm{x} \sum_{n,n^\prime} \hat{a}_{n^\prime}^\dagger
\rho_{n^\prime n} (\bm{x},t) \left[ \hat{\phi}_{\rm{mat}} (\bm{x},t) \right] \hat{a}_n,
\end{eqnarray}
with an inherent scalar potential operator for matter,
\begin{eqnarray}
\hat{\phi}_{\rm{mat}} (\bm{x},t) &=& \phi_{\rm ncl} (\bm{x},t) + \hat{\phi}_{\rm e-e} (\bm{x},t), \\
{
\hat{\phi}_{\rm e-e} (\bm{x},t)
}
&\equiv&
{
\frac{e}{4\pi\varepsilon_0} \int d\bm{x}^\prime
\frac{\hat{\Psi}^\dagger (\bm{x}^\prime,t) \hat{\Psi} (\bm{x}^\prime,t)}{|\bm{x} - \bm{x}^\prime|}.
}
\end{eqnarray}
Note that, in the absence of $\bm{A}$ and $\phi_{\rm ext}$, a material is electrically neutral,
and $\left\langle \hat{\phi}_{\rm{mat}} (\bm{x},t) \right\rangle_0 = 0$.
Here, we take the basis for the static state without external fields and consider $\hat{H}_0$ in
Eq.\ (\ref{eq:2ndH_0}) as a non-perturbative Hamiltonian.
When one applies the electromagnetic field of $\bm{A} (\bm{x},t)$ and $\phi_{\rm ext} (\bm{x},t)$ to the material,
a polarized charge is induced.
It gives an additional interaction between the induced polarized charges,
\begin{eqnarray}
\hat{H}_{\rm p-p}
&\simeq& \int d\bm{x} \frac{\hat{\rho} (\bm{x},t) - \rho_0 (\bm{x})}{4\pi\varepsilon_0}
\int d\bm{x}^\prime \frac{\hat{\rho} (\bm{x}^\prime,t) - \rho_0 (\bm{x}^\prime)}{|\bm{x} - \bm{x}^\prime|}
\nonumber \\
&=&
{
\int d\bm{x} \delta \hat{\rho} (\bm{x},t) \hat{\phi}_{\rm pol} (\bm{x},t).
}
\label{eq:Hpp2nd}
\end{eqnarray}
Here, $\rho_0 (\bm{x}) = \left\langle \hat{\rho} (\bm{x},t) \right\rangle_0$ is the average charge density from the electrons
[$\hat{\rho} (\bm{x},t) \equiv e \hat{\Psi}^\dagger (\bm{x},t) \hat{\Psi} (\bm{x},t)$].
Thus, $\delta \hat{\rho} (\bm{x},t) = \hat{\rho} (\bm{x},t) - \rho_0 (\bm{x})$ indicates a deviation from the static distribution.
The scalar potential due to the induced polarized charge is
\begin{equation}
{
\hat{\phi}_{\rm pol} (\bm{x},t)
= \int d\bm{x}^\prime
\frac{\delta \hat{\rho} (\bm{x}^\prime,t)}{4\pi\varepsilon_0 |\bm{x} - \bm{x}^\prime|}.
}
\label{eq:polscalar}
\end{equation}
Therefore, the L field consists of three elements: $\hat{\phi}_{\rm{mat}}$,
$\hat{\phi}_{\rm{pol}}$, and $\phi_{\rm{ext}}$, which are associated with
the electron--electron interaction in the static case, the induced charge densities, and the external field, respectively.

{
In this study, we divide the total Hamiltonian as $\hat{H} = \hat{H}_0 + \hat{H}_{\rm ind}$ with
\begin{equation}
\hat{H}_{\rm ind} = \hat{H}_{\rm int} + \hat{H}_{\rm ext} + \hat{H}_{\rm p-p}.
\label{eq:Hind}
\end{equation}
$\hat{H}_0$ describes the static system without any field irradiation, whereas
all induced effects by the external electric and magnetic fields are in $\hat{H}_{\rm ind}$.
Then, the incident and induced components are included in the perturbative term.
For the scalar potential of the induced polarized charge, we apply the mean field approximation,
$\hat{\Psi}^\dagger (\bm{x},t) \hat{\phi}_{\rm pol} (\bm{x},t) \hat{\Psi} (\bm{x},t)
\simeq \hat{\Psi}^\dagger (\bm{x},t) \hat{\Psi} (\bm{x},t) \phi_{\rm pol} (\bm{x},t) $.
For the external potential, $\int d\bm{x} \rho_0 (\bm{x}) \phi_{\rm ext} (\bm{x},t)$ corresponds to
the system energy shifts without excitation, as it is constant.
Subtracting this term from the Hamiltonian, we obtain
\begin{eqnarray}
\hat{H}_{\rm ind}^\prime
&=& \hat{H}_{\rm ind} - \left( \int d\bm{x} \rho_0 (\bm{x}) \phi_{\rm ext} (\bm{x},t) \right) \nonumber \\
&\simeq& - \int d\bm{x} \left[
\hat{\bm{j}} (\bm{x},t) \cdot \bm{A} (\bm{x},t) - \delta \hat{\rho} (\bm{x},t)
\Big( \phi_{\rm ext} (\bm{x},t) + \phi_{\rm pol} (\bm{x},t) \Big)
\right],
\label{eq:Hindp}
\end{eqnarray}
which describes the interaction between the matter and applied and induced EM fields for both the L and T components.
The L field is included in both the constitutive and Maxwell's equations.
However, we do not need to take care of a double count of the electron--electron interaction (see Appendix \ref{appx:anotherH}).
}

The susceptibility is obtained from the linear response theory~\cite{Kubo57}.
The susceptibility, field-induced charge density, and induced current density formulations depend on
the treatment of the non-perturbative and perturbative Hamiltonians.
Note that $\hat{H}_0$ includes neither a net charge nor current.
Therefore, for the formulation of the susceptibility based on
$\hat{H} = \hat{H}_0 + \hat{H}_{\rm ind}^\prime$ is evaluated by
the static states $|\mu \rangle$ for $\hat{H}_0$.
As this susceptibility is irrelevant to the induced fields, we can analyze
the T and L components of the field-induced charge and current in terms of
$\bm{A}$ (T field) and $\phi$ (L field).
A self-consistent relation between the induced charge and $\phi_{\rm{pol}}$ can describe
a plasmon (collective excitation) within the RPA~\cite{Prodan01}, as discussed in the following section.

We can omit $\delta$ of $\delta \hat{\rho} (\bm{x})$ in Eq.\ (\ref{eq:Hindp}) for simplicity.
In the following discussion, the charge density $\hat{\rho} (\bm{x})$ indicates the induced charge only.
This abbreviation does not change the continuous relation, as $\partial \rho_0 / \partial t = 0$.

\section{Four-vector representation for nonlocal susceptibility}
\label{sec:susceptibility}
The light--matter interaction $\hat{H}_{\rm ind}^\prime (t)$ in Eq.\ (\ref{eq:Hindp}) gives
a susceptibility to the external fields, $\bm{A} (\bm{x},t)$ and
$\phi_{\rm ind} (\bm{x},t) = \phi_{\rm ext} (\bm{x},t) + \phi_{\rm pol} (\bm{x},t)$ in linear response theory~\cite{Kubo57}.
Here, $\phi_{\rm ind}$ indicates a scalar potential caused by the field irradiation.
In the following, let us omit the subscript ``${\rm ind}$'' on $\hat{H}_{\rm ind}^\prime$ and $\phi_{\rm ind}$ for simplicity.
For the interaction Hamiltonian $\hat{H}^\prime (t)$, let us introduce the following four-vector representation:
\begin{eqnarray}
{
\bm{\mathcal{A}} (\bm{x},t)
}
&=&
{
\left( \begin{matrix}
\bm{A} (\bm{x},t) \\
- \frac{1}{c} \phi (\bm{x},t)
\end{matrix} \right),
}
\label{eq:4vecA} \\
\hat{\bm{\mathcal{J}}} (\bm{x},t)
&=& \left( \begin{matrix}
\hat{\bm{j}} (\bm{x},t) \\
c \hat{\rho} (\bm{x},t)
\end{matrix} \right),
\label{eq:4vecJ} \\
\hat{H}^\prime (t)
&=& - \int d\bm{x} \hat{\bm{\mathcal{J}}} (\bm{x},t) \cdot \bm{\mathcal{A}} (\bm{x},t).
\label{eq:4vecHind}
\end{eqnarray}
The time-dependence of $\hat{H}^\prime (t)$ in Eq.\ (\ref{eq:4vecHind})
is attributed to the external field $\bm{\mathcal{A}} (\bm{x},t)$ only, and $c$ is the light velocity in vacuum.
Note that we define the sign for the scalar potential component as negative in Eq.\ (\ref{eq:4vecA}) for a simpler formulation.

We assume a monochromatic field,
$\bm{\mathcal{A}} (\bm{x},t) = \bm{\mathcal{A}} (\bm{x};\omega) e^{-i\omega t}$.
The statistical average of the four-vector current $\hat{\bm{\mathcal{J}}} (\bm{x})$ gives
the susceptibility by the field $\bm{\mathcal{A}} (\bm{x};\omega)$:
\begin{eqnarray}
\left\langle \hat{\bm{\mathcal{J}}} (\bm{x}) \right\rangle (t)
&=&
{
\bm{\mathcal{J}} (\bm{x};\omega) e^{-i\omega t}
}
\nonumber \\
&=&
{
\bm{\mathcal{J}}_0 (\bm{x};\omega) e^{-i\omega t}
}
+ \int d\bm{x}^\prime \bar{\mathcal{X}} (\bm{x},\bm{x}^\prime ;\omega)
\cdot \bm{\mathcal{A}} (\bm{x}^\prime ;\omega) e^{-i\omega t},
\label{eq:cons4vec}
\end{eqnarray}
with a nonlocal susceptibility~\cite{book:cho1}
\begin{eqnarray}
\bar{\mathcal{X}} (\bm{x},\bm{x}^\prime ;\omega)
&\equiv& \sum_{\mu ,\nu}
\left[
f_{\nu \mu} \bm{\mathcal{J}}_{\mu \nu} (\bm{x}) \left( \bm{\mathcal{J}}_{\nu \mu} (\bm{x}^\prime) \right)^{\rm t}
+
h_{\nu \mu} \bm{\mathcal{J}}_{\nu \mu} (\bm{x}) \left( \bm{\mathcal{J}}_{\mu \nu} (\bm{x}^\prime) \right)^{\rm t}
\right] \nonumber \\
&=&
\left( \begin{matrix}
\bar{\chi}_{j j}             (\bm{x},\bm{x}^\prime ;\omega) & \bm{\chi}_{j \rho} (\bm{x},\bm{x}^\prime ;\omega) \\
\bm{\chi}_{\rho j}^{\rm \ t} (\bm{x},\bm{x}^\prime ;\omega) & \chi_{\rho \rho}   (\bm{x},\bm{x}^\prime ;\omega) 
\end{matrix} \right),
\label{eq:nonlocalsus4vec}
\end{eqnarray}
and
\begin{eqnarray}
\bm{\mathcal{J}}_{\nu \mu} (\bm{x})
&=& \left\langle \nu \left| \hat{\bm{\mathcal{J}}} (\bm{x}) \right| \mu \right\rangle,
\label{eq:Jmunu} \\
f_{\nu \mu} &=& \frac{\rho_{0,\mu}}{\hbar \omega_{\nu \mu} - \hbar \omega - i\gamma},
\label{eq:fmunu} \\
h_{\nu \mu} &=& \frac{\rho_{0,\mu}}{\hbar \omega_{\nu \mu} + \hbar \omega + i\gamma}.
\label{eq:hmunu}
\end{eqnarray}
The nonlocal susceptibility is a $4 \times 4$ matrix in the four-vector space and
$|\mu \rangle$ represents the electronic eigenstates of $\hat{H}_0$.
Thus, the susceptibility is affected by the nanostructure geometry via $|\mu \rangle$.
If the electronic system has translational symmetry, e.g., in the case of a bulk metal,
the nonlocality in the susceptibility is given only by a relative position, $\bar{\mathcal{X}} (\bm{x}-\bm{x}^\prime ;\omega)$.
Further, $\hbar \omega_{\nu \mu} = \varepsilon_\nu - \varepsilon_\mu$ is the energy difference between
the eigenenergies for $\hat{H}_0$, $\gamma$ is an infinitesimal positive value for the causality, and
$\rho_{0,\mu}$ in the numerators is an element of the density matrix at equilibrium, where
\begin{equation}
\rho_{0,\mu} = \langle \mu | \hat{\rho}_0 | \mu \rangle
= \frac{1}{Z_0} \langle \mu | e^{-\beta \hat{H}_0} | \mu \rangle
\label{eq:dm}
\end{equation}
with the partition function $Z_0 = {\rm Tr} \{ e^{-\beta \hat{H}_0} \}$.
{
The first term in the r.h.s. of Eq.\ (\ref{eq:cons4vec}) is
\begin{equation}
\bm{\mathcal{J}}_0 (\bm{x};\omega)
= \left( \begin{matrix}
\langle \hat{\bm{j}} (\bm{x};\omega) \rangle_0
\\
c \left( \langle \hat{\rho} (\bm{x}) \rangle_0 - \rho_0 (\bm{x}) \right)
\end{matrix} \right)
= \left( \begin{matrix}
- \frac{e}{m^*} \rho_0 (\bm{x}) \bm{A} (\bm{x};\omega)
\\
0
\end{matrix} \right)
\label{eq:AinJ}
\end{equation}
at equilibrium.
}
The elements of susceptibility in Eq.\ (\ref{eq:nonlocalsus4vec}) are related to each other and satisfy the continuous relation
$\bm{\nabla}_x \cdot \left\langle \hat{\bm{j}} (\bm{x}) \right\rangle -i\omega \left\langle \hat{\rho} (\bm{x}) \right\rangle =0$, where
\begin{eqnarray}
\bm{\nabla}_x \cdot \bar{\chi}_{jj} (\bm{x},\bm{x}^\prime ;\omega)
&=& i (\omega/c) \bm{\chi}_{\rho j}^{\rm \ t} (\bm{x},\bm{x}^\prime ;\omega), \\
\bm{\nabla}_x \cdot \bm{\chi}_{j \rho} (\bm{x},\bm{x}^\prime ;\omega)
&=& i (\omega/c) \chi_{\rho \rho} (\bm{x},\bm{x}^\prime ;\omega).
\end{eqnarray}

In the linear response, the $\bm{\mathcal{A}}$-dependence of the susceptibility
$\bar{\mathcal{X}} (\bm{x},\bm{x}^\prime ;\omega)$ cannot be discussed.
To evaluate the effect of the $\bm{A} (\bm{x},t)$ term in Eq.\ (\ref{eq:nonlocalsus4vec}),
the higher-order terms must be considered.

\section{Maxwell's equations for Coulomb gauge}
\label{sec:Maxwell}
Taking the Coulomb gauge, the vector and scalar potentials describe only the T and L components, respectively.
For the Maxwell's equations, we consider the potentials $\bm{\mathcal{A}} (\bm{x},t)$ induced by
the densities $\bm{\mathcal{J}} (\bm{x},t)$ via the Green's function.

The Maxwell's equations for the potentials, $\phi (\bm{x},t)$ and $\bm{A} (\bm{x},t)$, are expressed as
\begin{eqnarray}
\bm{\nabla}^2 \bm{A} (\bm{x},t) - \frac{1}{c^2} \frac{\partial^2 \bm{A} (\bm{x},t)}{\partial t^2}
- \frac{1}{c^2} \frac{\partial}{\partial t} \bm{\nabla} \phi (\bm{x},t)
&=& - \mu_0 \bm{j} (\bm{x},t),
\label{eq:CoulombJ}
\\
\bm{\nabla}^2 \phi (\bm{x},t)
&=& - \frac{\rho (\bm{x},t)}{\varepsilon_0}.
\label{eq:Coulombrho}
\end{eqnarray}
Note that the L component of the field is relevant only to $\rho (\bm{x},t)$, whereas $\bm{j} (\bm{x},t)$ generates
both of the L and T components.
Another description may also be considered (see Appendix~\ref{appx:LTMaxwell}).
Equations (\ref{eq:CoulombJ}) and (\ref{eq:Coulombrho}) are unified in the four-vector representation
\begin{equation}
\left( \begin{matrix}
\bm{\nabla}^2 - \frac{\partial^2}{\partial (ct)^2} &
{
\frac{\partial}{\partial (ct)} \bm{\nabla}
}
\\
0 &
{
- \bm{\nabla}^2
}
\end{matrix} \right)
\bm{\mathcal{A}} (\bm{x},t)
= - \mu_0
\bm{\mathcal{J}} (\bm{x},t),
\label{eq:Maxwell4vec}
\end{equation}
with the statistical average of the densities,
$\bm{\mathcal{J}} (\bm{x},t) = \left\langle \hat{\bm{\mathcal{J}}} (\bm{x}) \right\rangle (t)$.
For the Fourier component,
$\bm{\mathcal{O}} (\bm{x},t) = \bm{\mathcal{O}} (\bm{x} ;\omega) e^{-i\omega t}$
($\bm{\mathcal{O}} = \bm{\mathcal{A}},\bm{\mathcal{J}}$),
the Maxwell's equation is written as
\begin{equation}
\bar{\mathcal{D}} (\bm{x} ;\omega) \bm{\mathcal{A}} (\bm{x} ;\omega)
= - \mu_0 \bm{\mathcal{J}} (\bm{x} ;\omega),
\label{eq:Momega4vec}
\end{equation}
with the $4 \times 4$ differential matrix
\begin{equation}
\bar{\mathcal{D}} (\bm{x} ;\omega)
\equiv
\left( \begin{matrix}
\bm{\nabla}^2 + \frac{\omega^2}{c^2} &
{
-i \frac{\omega}{c} \bm{\nabla}
}
\\
0 &
{
- \bm{\nabla}^2
}
\end{matrix} \right).
\label{eq:D4vec}
\end{equation}

From Eqs.\ (\ref{eq:cons4vec}) and (\ref{eq:Momega4vec}),
we can construct a self-consistent relation between $\bm{\mathcal{J}}$ and $\bm{\mathcal{A}}$.
For the self-consistent equation, let us obtain the formal solution of
the Maxwell's equation (\ref{eq:Momega4vec}) using the Green's function
\begin{equation}
\bm{\mathcal{A}} (\bm{x} ;\omega)
= \bm{\mathcal{A}}_0 (\bm{x} ;\omega)
- \mu_0 \int d\bm{x}^\prime \bar{\mathcal{G}} (\bm{x},\bm{x}^\prime ;\omega)
\bm{\mathcal{J}} (\bm{x}^\prime ;\omega).
\label{eq:formalA4vec}
\end{equation}
The first term in Eq.\ (\ref{eq:formalA4vec}) corresponds to an ``incident'' field satisfying
\begin{equation}
\bar{\mathcal{D}} (\bm{x} ;\omega) \bm{\mathcal{A}}_0 (\bm{x} ;\omega) = 0.
\label{eq:A04vec}
\end{equation}
The Green's function is given as
\begin{equation}
\bar{\mathcal{G}} (\bm{x},\bm{x}^\prime ;\omega)
=
- \frac{1}{4\pi}
\left( \begin{matrix}
\frac{ \ e^{i\frac{\omega}{c}|\bm{x} - \bm{x}^\prime|} }{|\bm{x} - \bm{x}^\prime|} \bar{1} \ &
\frac{-i\omega}{4\pi c} \int d\bm{x}'' \frac{e^{i\frac{\omega}{c} |\bm{x} - \bm{x}''|}}{|\bm{x} - \bm{x}''|}
\frac{\bm{x}'' - \bm{x}^\prime}{|\bm{x}'' - \bm{x}^\prime|^3} \\
 0 &
{
- \frac{1}{|\bm{x} - \bm{x}^\prime|}
}
\end{matrix} \right).
\label{eq:solGreenMat}
\end{equation}
A detailed derivation of $\bar{\mathcal{G}}$ in Eq.\ (\ref{eq:solGreenMat}) is presented in Appendix \ref{appx:Green}.
This Green's function provides a scheme for field (potential) generation due to
the electronic excitations (current and charge densities).

\section{Self-consistent equation}
\label{sec:SelfLT}
In previous sections, we derived the constitutive equation (\ref{eq:cons4vec}) with
the nonlocal susceptibility (\ref{eq:nonlocalsus4vec}) and the solution of
the Maxwell's equations (\ref{eq:formalA4vec}) with Green's function (\ref{eq:solGreenMat}) in a four-vector representation.
These expressions form a self-consistent relation.

\subsection{$\bm{x}$-space representation}
\label{sec:x-rep}
Let us first formulate the self-consistent equation in terms of
$\bm{\mathcal{A}} (\bm{x};\omega)$ and $\bm{\mathcal{J}} (\bm{x};\omega)$.
Substitution of the constitutive equation into the solution of Maxwell's equation yields
{
the self-consistent equation for the four-vector potential $\bm{\mathcal{A}} (\bm{x};\omega)$,
\begin{eqnarray}
\bm{\mathcal{A}} (\bm{x};\omega)
&=&
\bm{\mathcal{A}}_0 (\bm{x};\omega)
- \mu_0 \int d\bm{x}' \bar{\mathcal{G}}(\bm{x},\bm{x}';\omega) \bm{\mathcal{J}}_0 (\bm{x}';\omega)
\nonumber \\
& & \hspace{20mm}
- \mu_0 \int d\bm{x}' \int d\bm{x}'' \bar{\mathcal{G}} (\bm{x},\bm{x}';\omega)
\bar{\mathcal{X}} (\bm{x}',\bm{x}'';\omega) \bm{\mathcal{A}} (\bm{x}'';\omega).
\label{eq:selfA}
\end{eqnarray}
By applying the separable integral kernel $\delta (\bm{x}-\bm{x}') = \sum_m \varphi_m^\ast (\bm{x}) \varphi_m (\bm{x}')$ for
the vector potential term (\ref{eq:AinJ}) in the current, we expand the delta function as
\begin{eqnarray}
(-\mu_0) \bm{\mathcal{J}}_0 (\bm{x};\omega)
&=&
\frac{\mu_0 e^2 n_0}{m^*} \int_V d\bm{x}^\prime \delta (\bm{x}-\bm{x}^\prime)
\left( \begin{matrix}
\bar{1} && \\
  &&  0
\end{matrix} \right)
\bm{\mathcal{A}} (\bm{x}^\prime;\omega)
\nonumber \\
&=&
\left( \frac{\omega_{\rm p}}{c} \right)^2 \sum_m \sum_{\alpha =x,y,z} \int_V d\bm{x}^\prime
\left( \varphi_m^\ast (\bm{x}) \bm{e}_\alpha \right)
\left( \varphi_m (\bm{x}^\prime) \bm{e}_\alpha \right)^{\rm t} \bm{\mathcal{A}} (\bm{x}';\omega).
\label{eq:exvectorterm}
\end{eqnarray}
Here, $n_0 = \rho_0/e$ is the electron density and $\omega_{\rm p} = \sqrt{e^2 n_0/(\varepsilon_0 m^*)}$
is the plasma frequency in bulk.
$\bm{e}_\alpha$ means a unit vector in the $\alpha$ direction.
The integral is applied only inside the nanostructure.
}

Since the nonlocal susceptibility $\bar{\mathcal{X}} (\bm{x}',\bm{x}'' ;\omega)$ shown in
Eq.\ (\ref{eq:nonlocalsus4vec}) is a separable integral kernel,
by multiplying Eq.\ (\ref{eq:selfA}) by $\left( \bm{\mathcal{J}}_{\nu' \mu'} (\bm{x}) \right)^{\rm t}$,
$\left( \bm{\mathcal{J}}_{\mu' \nu'} (\bm{x}) \right)^{\rm t}$,
{and $\left( \varphi_{m'} (\bm{x}) \bm{e}_\beta^{\rm t} \right)$
}
from the left and integrating with respect to $\bm{x}$, we formulate the matrix form of the self-consistent equation as
{
\begin{equation}
\left[ \bar{\Xi} (\omega) \right]
\left( \begin{matrix}
\bm{X}^{(-)} (\omega) \\
\bm{X}^{(+)} (\omega)
\end{matrix} \right)
=
\left( \begin{matrix}
\bm{Y}^{\prime (0,-)} (\omega) \\
\bm{Y}^{\prime (0,+)} (\omega)
\end{matrix} \right)
\label{eq:exSCmat}
\end{equation}
with
\begin{equation}
\bar{\Xi} (\omega)
=
\left( \begin{matrix}
\hbar \bar{\Omega} - (\hbar \omega + i\gamma) \bar{1} & 
\\
&
\hbar \bar{\Omega} + (\hbar \omega + i\gamma) \bar{1} \\
\end{matrix} \right)
+
\left( \begin{matrix}
\bar{K}^\prime (\omega) & \bar{L}^\prime (\omega) \\
\bar{M}^\prime (\omega) & \bar{N}^\prime (\omega)
\end{matrix} \right).
\label{eq:exSCmatXi}
\end{equation}
The vectors
\begin{equation}
\left( \begin{matrix}
\bm{Y}^{\prime (0,-)} (\omega) \\
\bm{Y}^{\prime (0,+)} (\omega)
\end{matrix} \right)
=
\left( \begin{matrix}
\bm{Y}^{(0,-)} (\omega) \\
\bm{Y}^{(0,+)} (\omega)
\end{matrix} \right)
+
\left( \begin{matrix}
\bar{U} (\omega) \frac{1}{\bar{1} - \bar{R} (\omega)} \bm{Y}^{(A)} (\omega) \\
\bar{V} (\omega) \frac{1}{\bar{1} - \bar{R} (\omega)} \bm{Y}^{(A)} (\omega)
\end{matrix} \right),
\label{eq:exSCmatY}
\end{equation}
correspond to an incident field.
The detail derivation and formulation are described in Appendix \ref{appx:SCmatrix}.
$\bar{\Omega} = {\rm diag} (\bar{\Omega}_0, \bar{\Omega}_1, \cdots ,\bar{\Omega}_N)$ is
a diagonal matrix with $\bar{\Omega}_\mu = {\rm diag} (\omega_{0\mu},\omega_{1\mu},\omega_{2\mu}\cdots,\omega_{N\mu})$
for the excitation energy $\hbar \omega_{\nu \mu} = \varepsilon_\nu - \varepsilon_\mu$.
$\gamma$ is an infinitesimal value for the causality.
The matrices 
\begin{eqnarray}
\bar{K}^\prime (\omega)
&=&
\bar{K} (\omega) + \bar{U} (\omega) \frac{1}{\bar{1} - \bar{R} (\omega)} \bar{S} (\omega), \\
\bar{L}^\prime (\omega)
&=&
\bar{L} (\omega) + \bar{U} (\omega) \frac{1}{\bar{1} - \bar{R} (\omega)} \bar{T} (\omega), \\
\bar{M}^\prime (\omega)
&=&
\bar{M} (\omega) + \bar{V} (\omega) \frac{1}{\bar{1} - \bar{R} (\omega)} \bar{S} (\omega), \\
\bar{N}^\prime (\omega)
&=&
\bar{N} (\omega) + \bar{V} (\omega) \frac{1}{\bar{1} - \bar{R} (\omega)} \bar{T} (\omega).
\end{eqnarray}
describe the radiative correction or the correlation between the excited charge and current densities by
the longitudinal and transverse fields.
Their matrix and vector components are defined as
\begin{eqnarray}
K_{\nu' \mu' ,\mu \nu} (\omega)
&=&
\mu_0 \rho_{0,\mu} \int d\bm{x} \int d\bm{x}' \left( \bm{\mathcal{J}}_{\nu' \mu'} (\bm{x}) \right)^{\rm t}
\bar{\mathcal{G}} (\bm{x},\bm{x}' ;\omega) \bm{\mathcal{J}}_{\mu \nu}(\bm{x}'),
\label{eq:KJGJ} \\
L_{\nu' \mu' ,\nu \mu} (\omega)
&=&
\mu_0 \rho_{0,\mu} \int d\bm{x} \int d\bm{x}' \left( \bm{\mathcal{J}}_{\nu' \mu'} (\bm{x}) \right)^{\rm t}
\bar{\mathcal{G}} (\bm{x},\bm{x}' ;\omega) \bm{\mathcal{J}}_{\nu \mu}(\bm{x}'),
\label{eq:LJGJ} \\
M_{\mu' \nu' ,\mu \nu} (\omega)
&=&
\mu_0 \rho_{0,\mu} \int d\bm{x} \int d\bm{x}'
\left( \bm{\mathcal{J}}_{\mu' \nu'} (\bm{x}) \right)^{\rm t}
\bar{\mathcal{G}}(\bm{x},\bm{x}' ;\omega) \bm{\mathcal{J}}_{\mu \nu} (\bm{x}'),
\label{eq:MJGJ} \\
N_{\mu' \nu' ,\nu \mu} (\omega)
&=&
\mu_0 \rho_{0,\mu} \int d\bm{x} \int d\bm{x}'
\left( \bm{\mathcal{J}}_{\mu' \nu'} (\bm{x}) \right)^{\rm t}
\bar{\mathcal{G}}(\bm{x},\bm{x}' ;\omega) \bm{\mathcal{J}}_{\nu \mu} (\bm{x}'),
\label{eq:NJGJ} \\
R_{m' \beta ,m \alpha} (\omega)
&=&
\left( \frac{\omega_{\rm p}}{c} \right)^2
\int d\bm{x} \int d\bm{x}' \left( \varphi_{m'} (\bm{x}) \bm{e}_\beta \right)^{\rm t}
\bar{\mathcal{G}} (\bm{x},\bm{x}';\omega)
\left( \varphi_m^\ast (\bm{x}') \bm{e}_\alpha \right),
\label{eq:exSFR} \\
U_{\nu'\mu',m \alpha} (\omega)
&=&
\left( \frac{\omega_{\rm p}}{c} \right)^2
\int d\bm{x} \int d\bm{x}' \left( \bm{\mathcal{J}}_{\nu' \mu'}(\bm{x}) \right)^{\rm t}
\bar{\mathcal{G}} (\bm{x},\bm{x}';\omega)
\left( \varphi_m^\ast (\bm{x}') \bm{e}_\alpha \right),
\label{eq:exSFU} \\
V_{\mu'\nu',m \alpha} (\omega)
&=&
\left( \frac{\omega_{\rm p}}{c} \right)^2
\int d\bm{x} \int d\bm{x}' \left( \bm{\mathcal{J}}_{\mu' \nu'}(\bm{x}) \right)^{\rm t}
\bar{\mathcal{G}} (\bm{x},\bm{x}';\omega)
\left( \varphi_m^\ast (\bm{x}') \bm{e}_\alpha \right),
\label{eq:exSFV} \\
S_{m' \beta ,\mu \nu} (\omega)
&=&
\mu_0 \rho_{0,\mu} \int d\bm{x} \int d\bm{x}' \left( \varphi_{m'} (\bm{x}) \bm{e}_\beta \right)^{\rm t}
\bar{\mathcal{G}} (\bm{x},\bm{x}';\omega) \bm{\mathcal{J}}_{\mu \nu} (\bm{x}'),
\label{eq:exSFS} \\
T_{m' \beta ,\nu \mu} (\omega)
&=&
\mu_0 \rho_{0,\mu} \int d\bm{x} \int d\bm{x}' \left( \varphi_{m'} (\bm{x}) \bm{e}_\beta \right)^{\rm t}
\bar{\mathcal{G}} (\bm{x},\bm{x}';\omega) \bm{\mathcal{J}}_{\nu \mu} (\bm{x}')
\label{eq:exSFT}
\end{eqnarray}
and
\begin{eqnarray}
X_{\nu \mu}^{(-)} (\omega)
&=&
\frac{1}{\hbar \omega_{\nu \mu} - \hbar \omega - i\gamma}
\int d\bm{x} \left( \bm{\mathcal{J}}_{\nu \mu} (\bm{x}) \right)^{\rm t}
\bm{\mathcal{A}} (\bm{x} ;\omega),
\label{eq:X-JA} \\
X_{\mu \nu}^{(+)} (\omega)
&=&
\frac{1}{\hbar \omega_{\nu \mu} + \hbar \omega + i\gamma}
\int d\bm{x} \left( \bm{\mathcal{J}}_{\mu \nu} (\bm{x}) \right)^{\rm t}
\bm{\mathcal{A}} (\bm{x} ;\omega),
\label{eq:X+JA} \\
X_{m \alpha}^{(A)} (\omega)
&=&
\int d\bm{x} \left( \varphi_m (\bm{x}) \bm{e}_\alpha \right)^{\rm t} \bm{\mathcal{A}} (\bm{x};\omega),
\label{eq:exSFXA} \\
Y^{\rm (0)}_{\nu' \mu'} (\omega)
&=& \int d\bm{x} \left( \bm{\mathcal{J}}_{\nu' \mu'} (\bm{x}) \right)^{\rm t}
\bm{\mathcal{A}}_0 (\bm{x} ;\omega),
\label{eq:Y0JA} \\
Y_{m \alpha}^{(A)} (\omega)
&=&
\int d\bm{x} \left( \varphi_m (\bm{x}) \bm{e}_\alpha \right)^{\rm t} \bm{\mathcal{A}}_0 (\bm{x};\omega).
\label{eq:exSFYA}
\end{eqnarray}
}
{
The matrix $\bar{\Xi} (\omega)$ describes coupling between the electron--hole excitations mediated by
both longitudinal and transverse electromagnetic fields, which forms the collective excitation.
The individual electron--hole excitations are followed by $\bar{\Omega}$, whereas
the field-mediated coupling, namely the collective excitation, is in $\bar{K}^\prime (\omega)$, $\bar{L}^\prime (\omega)$,
$\bar{M}^\prime (\omega)$, and $\bar{N}^\prime (\omega)$.
Due to coupling, the eigenvalues of $\bar{\Xi} (\omega)$ become complex and provide zero points in
the complex $\omega = \omega_{\rm r} + i \omega_{\rm i}$ plane.
Its real components provide the excitation spectrum, and the imaginary components means the radiative width.
The vectors $\bm{X}^{(\mp)} (\omega)$ and ${\bm{Y}^\prime}^{(0,\mp)} (\omega)$ in Eq.\ (\ref{eq:exSCmat})
mean the coefficients of induced four-vector current and incident four-vector field, respectively.
Therefore, the matrix $\bar{\Xi} (\omega)$ exhibits the electronic properties of nanostructures,
whereas the vector $\bm{X}^{(\mp)} (\omega)$ provides the output charge and current densities.
}

At zero temperature, i.e., $T=0$, $\rho_{0,\mu} = \delta_{0,\mu}$ in $f_{\nu \mu}$ and $h_{\nu \mu}$
defined in Eqs.\ (\ref{eq:fmunu}) and (\ref{eq:hmunu}), respectively.
Then, the energy differences $\hbar \omega_{\nu 0}$ in the denominators of
$X_{\nu 0}^{(-)}$ and $X_{0 \nu}^{(+)}$ are positive.
For the matrix form of the self-consistent equation (\ref{eq:exSCmat}),
only $\bm{\mathcal{J}}_{\nu 0}$ and $\bm{\mathcal{J}}_{0 \nu}$ should be considered.
Therefore, the matrix elements are reduced for
$K_{\nu' 0,0 \nu}$, $L_{\nu' 0,\nu 0}$, $M_{0 \nu' ,0 \nu}$, and $N_{0 \nu' ,\nu 0}$.

The matrix form of the self-consistent equation (\ref{eq:exSCmat}) has an important advantage.
The matrix $\bar{\Xi} (\omega)$ with the matrices $\bar{K}$, $\bar{L}$, $\bar{M}$, and $\bar{N}$
consisting of $\bar{\mathcal{G}}$ and $\left\langle \nu \left| \hat{\bm{\mathcal{J}}} \right| \mu \right\rangle$
gives the eigenmodes of the system coupled with the radiative field.
The L and T fields are in $\bar{\mathcal{G}}$ and the properties of the nonlocal response are given by $| \mu \rangle$.
From the real and imaginary components of the eigenvalues of $\bar{\Xi} (\omega)$,
the properties of the collective and individual excitations are evaluated.
Then, the current and charge densities are discussed in terms of the plasmons and SPEs.

\subsection{$\bm{k}$-space representation}
\label{sec:k-rep}
For a metallic structure having some symmetry, the Fourier transformation of the fields and densities is useful for the evaluation.
In that case, we consider the $\bm{k}$-space representation for the self-consistent equation.
The fields and densities in the $\bm{k}$-representation are given as
\begin{equation}
\tilde{\bm{\mathcal{O}}} (\bm{k} ;\omega) = \frac{1}{(2\pi)^{\frac{3}{2}}} \int d\bm{x}
\bm{\mathcal{O}} (\bm{x} ;\omega) e^{-i\bm{k} \cdot \bm{x}},
\hspace{5mm}
(\bm{\mathcal{O}} = \bm{\mathcal{A}}, \bm{\mathcal{J}}).
\label{eq:4vecFourier}
\end{equation}
The Maxwell's equations (\ref{eq:Momega4vec}) and their solutions (\ref{eq:formalA4vec}) are
also transferred to the $\bm{k}$-representation, with the latter being expressed as
\begin{equation}
\tilde{\bm{\mathcal{A}}} (\bm{k} ;\omega)
=
\tilde{\bm{\mathcal{A}}}_0 (\bm{k} ;\omega)
- \mu_0 \bar{\mathcal{G}}_{\bm{k}} (\omega)
\tilde{\bm{\mathcal{J}}} (\bm{k} ;\omega).
\label{eq:formalA4vec_k}
\end{equation}
Here, the Green's function is given as
\begin{equation}
\bar{\mathcal{G}}_{\bm{k}} (\omega) =
\left( \begin{matrix}
\frac{1}{-\bm{k}^2 + \omega^2/c^2} \bar{1} &
-\frac{1}{\bm{k}^2} \frac{1}{-\bm{k}^2 + \omega^2/c^2} \frac{\omega}{c} \bm{k} \\
0 &
{
\frac{1}{\bm{k}^2}
}
\end{matrix} \right).
\label{eq:solGreenMat_k}
\end{equation}
The incident term satisfies
\begin{equation}
\left( \begin{matrix}
\left( -\bm{k}^2 + \frac{\omega^2}{c^2} \right) \bar{1} &
{
\frac{\omega}{c} \bm{k}
}
\\
0 &
{
\bm{k}^2
}
\end{matrix} \right)
\tilde{\bm{\mathcal{A}}}_0 (\bm{k} ;\omega) = 0.
\label{eq:A04vec_k}
\end{equation}
Note that the solution of the Maxwell's equations contains no integral.

The constitutive equation becomes
\begin{equation}
\tilde{\bm{\mathcal{J}}} (\bm{k} ;\omega)
= \tilde{\bm{\mathcal{J}}}_0 (\bm{k} ;\omega) + \int d\bm{k}^\prime \bar{\mathcal{X}}_{\bm{k},\bm{k}^\prime} (\omega)
\tilde{\bm{\mathcal{A}}} (\bm{k}^\prime ;\omega),
\label{eq:cons4vec_k}
\end{equation}
with the nonlocal susceptibility
\begin{equation}
\bar{\mathcal{X}}_{\bm{k},\bm{k}^\prime} (\omega)
= \sum_{\mu ,\nu}
\left[
f_{\nu \mu} \tilde{\bm{\mathcal{J}}}_{\mu \nu} (\bm{k})
\left( \tilde{\bm{\mathcal{J}}}_{\nu \mu} (-\bm{k}^\prime) \right)^{\rm t}
+
h_{\nu \mu} \tilde{\bm{\mathcal{J}}}_{\nu \mu} (\bm{k})
\left( \tilde{\bm{\mathcal{J}}}_{\mu \nu} (-\bm{k}^\prime) \right)^{\rm t}
\right].
\label{eq:nonlocalsus4vec_k}
\end{equation}

Equations (\ref{eq:formalA4vec_k}) and (\ref{eq:cons4vec_k}) form the self-consistent equation:
{
\begin{equation}
\tilde{\bm{\mathcal{A}}} (\bm{k};\omega)
=
\tilde{\bm{\mathcal{A}}}_0 (\bm{k};\omega)
- \mu_0 \bar{\mathcal{G}}_{\bm{k}} (\omega) \tilde{\bm{\mathcal{J}}}_0 (\bm{k};\omega)
- \mu_0 \int d\bm{k}^\prime \bar{\mathcal{G}}_{\bm{k}} (\omega) \bar{\mathcal{X}}_{\bm{k},\bm{k}^\prime} (\omega)
\tilde{\bm{\mathcal{A}}} (\bm{k}^\prime;\omega).
\label{eq:SCk}
\end{equation}
}
The matrix form of the self-consistent equation in the $\bm{k}$-representation is obtained in the same manner as
that of the previous subsection.
It is exactly identical with Eq.\ (\ref{eq:exSCmat}), see Appendix \ref{appx:SCmatrix}.
The matrix elements in the $\bm{k}$-representation are , e.g., as
\begin{eqnarray}
\tilde{K}_{\nu' \mu' ,\mu \nu} (\omega)
&=&
\mu_0 \rho_{0,\mu} \int d\bm{k} \left( \tilde{\bm{\mathcal{J}}}_{\nu' \mu'} (-\bm{k}) \right)^{\rm t}
\bar{\mathcal{G}}_{\bm{k}} (\omega) \tilde{\bm{\mathcal{J}}}_{\mu \nu}(\bm{k}),
\label{eq:KJGJ_k} \\
\tilde{X}_{\nu \mu}^{(-)} (\omega)
&=&
\frac{1}{\hbar \omega_{\nu \mu} - \hbar \omega - i\gamma}
\int d\bm{k} \left( \tilde{\bm{\mathcal{J}}}_{\nu \mu} (-\bm{k}) \right)^{\rm t}
\tilde{\bm{\mathcal{A}}} (\bm{k} ;\omega).
\label{eq:X-JA_k}
\end{eqnarray}
The Fourier transformation demonstrates that these elements are equivalent to those of the $\bm{x}$-representation,
$\tilde{K}_{\nu' \mu' ,\mu \nu} (\omega) = K_{\nu' \mu' ,\mu \nu} (\omega)$,
$\tilde{X}_{\nu \mu}^{(-)} (\omega) = X_{\nu \mu}^{(-)} (\omega)$, etc.
Therefore, the matrix element evaluation can be performed in either the $\bm{x}$- or $\bm{k}$-representations depending on the case.

It is worth noting that, in $\left( \tilde{\bm{\mathcal{J}}}_{\nu' \mu'} (-\bm{k}) \right)^{\rm t}
\bar{\mathcal{G}}_{\bm{k}} (\omega) \tilde{\bm{\mathcal{J}}}_{\mu \nu}(\bm{k})$, a component between the induced charge densities,
\begin{equation}
K_{0,\nu' \mu' ,\mu \nu} = \mu_0 \rho_{0,\mu} \int d\bm{k} 
c\tilde{\rho}_{\nu' \mu'} (-\bm{k}) \frac{1}{\bm{k}^2} c\tilde{\rho}_{\mu \nu} (\bm{k}),
\end{equation}
describes the Coulomb interaction, which arises when an external field is applied.
{
If the contribution of the current densities is negligible, the self-consistent equation (\ref{eq:exSCmat}) is reduced as
\begin{equation}
\left[ \bar{\Xi}_0 (\omega) \right]
\left( \begin{matrix}
\bm{X}^{(-)} (\omega) \\
\bm{X}^{(+)} (\omega)
\end{matrix} \right)
=
\left( \begin{matrix}
\bm{Y}^{(0,-)} (\omega) \\
\bm{Y}^{(0,+)} (\omega)
\end{matrix} \right)
\label{eq:SCmatk0}
\end{equation}
with
\begin{equation}
\bar{\Xi}_0 (\omega)
=
\left( \begin{matrix}
\hbar \bar{\Omega} - (\hbar \omega + i\gamma) \bar{1} &  \\
                                                      & \hbar \bar{\Omega} + (\hbar \omega + i\gamma) \bar{1} \\
\end{matrix} \right)
+
\left( \begin{matrix}
\bar{K}_0 & \bar{K}_0 \\
\bar{K}_0 & \bar{K}_0
\end{matrix} \right).
\label{eq:exSCmatXik0}
\end{equation}
The eigenvalues of the matrix $\bar{\Xi}_0 (\omega)$ generate the spectrum of
the individual (electron-hole) excitation and the collective (plasmon) excitation.
It follows
\begin{equation}
{\rm det} \left\{ \hbar^2 \bar{\Omega}^2 - (\hbar \omega + i\gamma)^2 +
\hbar \bar{K}_0 \bar{\Omega} + \hbar \bar{\Omega} \bar{K}_0 \right\} = 0.
\end{equation}
A rough estimate of the spectrum is
\begin{equation}
1 \sim K_0 \frac{\hbar \omega_{\nu \mu}}{ (\hbar \omega + i\gamma)^2 - (\hbar \omega_{\nu \mu})^2}.
\end{equation}
Hence, we find the spectrum at $\omega \sim \omega_{\nu \mu}$ for the electron-hole excitation and
$\omega \gg \omega_{\nu \mu}$ for the plasmon excitation
In a free electron model for the basis $|\mu \rangle$ of $\hat{H}_0$, the plasmon spectrum should correspond to that in the RPA.
}

In addition to the $\tilde{\rho}_{\nu' \mu'} \left[ \bar{\mathcal{G}}_{\bm{k}} \right]_{\rho \rho} \tilde{\rho}_{\mu \nu}$ term,
$\tilde{K}_{\nu' \mu' ,\mu \nu}$ and the other matrix components have
$\tilde{\bm{j}}_{\nu' \mu'} \left[ \bar{\mathcal{G}}_{\bm{k}} \right]_{j \rho} \tilde{\rho}_{\mu \nu}$ and
$\tilde{\bm{j}}_{\nu' \mu'} \left[ \bar{\mathcal{G}}_{\bm{k}} \right]_{jj} \tilde{\bm{j}}_{\mu \nu}$ terms,
which have been neglected in many previous studies.
However, the modifications of the eigenstates and spectrum of the electronic system due to the coupled radiative field are
given by these components, as noted in the previous subsection.
The nanostructure shape determines the spectrum shift via the $| \mu \rangle$ in
$\tilde{\rho}_{\mu \nu}$ and $\tilde{\bm{j}}_{\mu \nu}$.
Therefore, analysis and exploration of an enhancement condition based on $\tilde{K}$, $\tilde{L}$, $\tilde{M}$, and $\tilde{N}$,
with $\tilde{\rho}_{\nu' \mu'} \left[ \bar{\mathcal{G}}_{\bm{k}} \right]_{\rho \rho} \tilde{\rho}_{\mu \nu}$,
$\tilde{\bm{j}}_{\nu' \mu'} \left[ \bar{\mathcal{G}}_{\bm{k}} \right]_{j \rho} \tilde{\rho}_{\mu \nu}$, and
$\tilde{\bm{j}}_{\nu' \mu'} \left[ \bar{\mathcal{G}}_{\bm{k}} \right]_{jj} \tilde{\bm{j}}_{\mu \nu}$ are important.

{
\section{Application example: Rectangular Nanorod}
\label{sec:RecRod}
To verify the formulation of self-consistent matrix equation (\ref{eq:exSCmat}) with Eq.\ (\ref{eq:exSCmatXi}) and
the feasibility of calculation, we apply the formulation to a single rectangular nanorod.
}

{
\subsection{Radiative correction matrix for rectangular nanorod}
\label{sec:XimatrixNR}
Electron and hole wavefunctions in a rectangular nanorod are given as
\begin{eqnarray}
\psi_{({\rm e}\mu) = (n_x,n_y,n_z)} (\bm{x})
&=& \sqrt{\frac{2}{L_x}} \sin \left( \frac{n_x \pi}{L_x} x \right)
    \sqrt{\frac{2}{L_y}} \sin \left( \frac{n_y \pi}{L_y} y \right)
    \sqrt{\frac{2}{L_z}} \sin \left( \frac{n_z \pi}{L_z} z \right),
\label{eq:wavefunc_e} \\
\psi_{({\rm h}\bar{\mu}) = (\bar{n}_x,\bar{n}_y,\bar{n}_z)} (\bm{x})
&=& \sqrt{\frac{2}{L_x}} \sin \left( \frac{\bar{n}_x \pi}{L_x} x \right)
    \sqrt{\frac{2}{L_y}} \sin \left( \frac{\bar{n}_y \pi}{L_y} y \right)
    \sqrt{\frac{2}{L_z}} \sin \left( \frac{\bar{n}_z \pi}{L_z} z \right)
\label{eq:wavefunc_h}
\end{eqnarray}
with $L_x$, $L_y$, and $L_z$ being the length of nanorod in the $x$, $y$, and $z$ directions, respectively.
This section aims to verify calculation feasibility.
We suppose the basis $|\mu =(\rm{e}\mu ,\rm{h}\bar{\mu}) \rangle$ for the Hamiltonian $\hat{H}_0$
are given by Eqs.\ (\ref{eq:wavefunc_e}) and (\ref{eq:wavefunc_h}).
Note that the electron and hole energies are $\varepsilon_{{\rm e}\mu} > \varepsilon_{\rm F}$ and
$\varepsilon_{{\rm h} \bar{\mu}} \le \varepsilon_{\rm F}$, respectively.
The excited charge and current densities at zero temperature are evaluated by the wavefunctions as
\begin{eqnarray}
\langle 0 | \hat{\rho} (\bm{x})| \mu \rangle
&=&
\rho_{0\mu} (\bm{x})
=
e \psi_{(\rm{h}\bar{\mu})} (\bm{x}) \psi_{({\rm e}\mu)} (\bm{x}),
\label{eq:rho_he}
\\
\langle 0 | \hat{\bm{j}} (\bm{x})| \mu \rangle
&=&
\bm{j}_{0\mu} (\bm{x})
=
-\frac{e\hbar}{2im^*} \left[
\left( \bm{\nabla} \psi_{(\rm{h}\bar{\mu})} (\bm{x}) \right)         \psi_{({\rm e}\mu)} (\bm{x})
-               \psi_{(\rm{h}\bar{\mu})} (\bm{x}) \left( \bm{\nabla} \psi_{({\rm e}\mu)} (\bm{x}) \right)
\right].
\label{eq:j_he}
\end{eqnarray}
Since the wavefunctions are real, we find $\rho_{\mu 0} (\bm{x}) = \rho_{0\mu} (\bm{x})$
and $\bm{j}_{\mu 0} (\bm{x}) = - \bm{j}_{\mu 0} (\bm{x})$.
}

{
By applying the Fourier transformation of the densities, the matrix elements of
$\bar{K}$, $\bar{L}$, $\bar{M}$, and $\bar{N}$ are evaluated as
\begin{eqnarray}
\tilde{K}_{\mu' 0,0\mu}
&=&
\mu_0 \int d\bm{k}
\left[
\left(\tilde{\bm{j}}_{\mu' 0}(-\bm{k}) \right)^{\rm t}
\left( \frac{1}{-\bm{k}^2 + (n_{\rm bg} \omega/c)^2} \right) \tilde{\bm{j}}_{0\mu}(\bm{k})
\right. \nonumber \\
& & \hspace{20mm}
+
\left(\tilde{\bm{j}}_{\mu' 0}(-\bm{k}) \right)^{\rm t}
\left( - \frac{1}{\bm{k}^2} \frac{n_{\rm bg} \omega/c}{- \bm{k}^2 + (n_{\rm bg} \omega/c)^2} \bm{k} \right)
(c/n_{\rm bg}) \tilde{\rho}_{0 \mu}(\bm{k})
\nonumber \\
& & \hspace{20mm} \left.
+
(c/n_{\rm bg}) \tilde{\rho}_{\mu' 0}(-\bm{k}) \left( \frac{1}{\bm{k}^2} \right) (c/n_{\rm bg}) \tilde{\rho}_{0 \mu}(\bm{k})
\right]
\nonumber \\
&=&
A_{\mu',\mu}^{(2)} + A_{\mu',\mu}^{(1)} + A_{\mu',\mu}^{(0)},
\label{eq:Kmu00mu} \\
\tilde{L}_{\mu' 0,\mu 0}
&=&
- A_{\mu',\mu}^{(2)} + A_{\mu',\mu}^{(1)} + A_{\mu',\mu}^{(0)},
\label{eq:Lmu0mu0} \\
\tilde{M}_{0 \mu' ,0 \mu}
&=&
- A_{\mu',\mu}^{(2)} - A_{\mu',\mu}^{(1)} + A_{\mu',\mu}^{(0)},
\label{eq:M0mu0mu} \\
\tilde{N}_{0 \mu',\mu 0}
&=&
A_{\mu',\mu}^{(2)} - A_{\mu',\mu}^{(1)} + A_{\mu',\mu}^{(0)}.
\label{eq:N0mumu0}
\end{eqnarray}
Here, $n_{\rm bg} \omega/c$ means the wavenumber of light in the nanostructure and environment with the refractive index $n_{\rm bg}$,
for which the background dielectric constant $\varepsilon_{\rm bg}$ is introduced phenomenologically,
$n_{\rm bg} = \sqrt{\varepsilon_{\rm bg} /\varepsilon_0}$.
The light velocity $c$ in the Maxwell's equations (\ref{eq:CoulombJ}) and (\ref{eq:Coulombrho})
[or Eqs.\ (\ref{eq:Maxwell4vec}) with the four-vector definition in Eqs.\ (\ref{eq:4vecA}) and (\ref{eq:4vecJ})] is replaced by $c/n_{\rm bg}$.
The back ground refractive index $n_{\rm bg}$ modulates the wavenumber of light,
which enlarges the current--current interaction $A_{\mu',\mu}^{(2)}$ and the current--charge interaction $A_{\mu',\mu}^{(1)}$
and reduces effectively the charge--charge interaction $A_{\mu',\mu}^{(0)}$ by the screening effect.
}

{
For the separable integral kernel in the elements of matrices
$\bar{S}$, $\bar{T}$, $\bar{U}$, $\bar{V}$, and $\bar{R}$, we use
\begin{equation}
\varphi_m (\bm{x}) =
\sqrt{\frac{2}{L_x}} \sin \left( \frac{m_x \pi}{L_x} x \right)
\sqrt{\frac{2}{L_y}} \sin \left( \frac{m_y \pi}{L_y} y \right)
\sqrt{\frac{2}{L_z}} \sin \left( \frac{m_z \pi}{L_z} z \right)
\label{eq:phi_m}
\end{equation}
with $m = (m_x,m_y,m_z)$.
Its Fourier  transferred function satisfies
$\sum_m \tilde{\varphi}_m^* (\bm{k}) \tilde{\varphi}_m (-\bm{k}') = \delta (\bm{k} - \bm{k}')$.
Then, the matrix elements are
\begin{eqnarray}
\tilde{S}_{m'\beta,0 \mu}
&=&
\mu_0 \int d\bm{k} \tilde{\varphi}_{m'} (-\bm{k})
\left(\frac{1}{- \bm{k}^2 + (n_{\rm bg} \omega/c)^2} \right) \tilde{j}^{(\beta)}_{0\mu} (\bm{k})
\nonumber \\
& & \hspace{15mm}
+
\mu_0 \int d\bm{k} \tilde{\varphi}_{m'} (-\bm{k})
\left(-\frac{1}{\bm{k}^2} \frac{n_{\rm bg} \omega/c}{- \bm{k}^2 + (n_{\rm bg} \omega/c)^2} k_\beta \right)
(c/n_{\rm bg}) \tilde{\rho}_{0\mu} (\bm{k})
\nonumber \\
&=&
B_{m'\beta,\mu}^{(2)} + B_{m'\beta,\mu}^{(1)},
\label{eq:Smbmu0} \\
\tilde{T}_{m'\beta,\mu 0}
&=&
- B_{m'\beta,\mu}^{(2)} + B_{m'\beta,\mu}^{(1)},
\label{eq:Tmb0mu} \\
\tilde{U}_{\mu' 0,m\alpha}
&=&
\left( \frac{\omega_{\rm p}^\prime}{c/n_{\rm bg}} \right)^2 \int d\bm{k} \tilde{j}^{(\alpha)}_{\mu'0} (-\bm{k})
\left( \frac{1}{- \bm{k}^2 + (n_{\rm bg} \omega/c)^2} \right) \tilde{\varphi}_m^\ast (\bm{k})
\nonumber \\
&=&
- \frac{1}{\mu_0} \left( \frac{\omega_{\rm p}^\prime}{c/n_{\rm bg}} \right)^2 B_{m\alpha,\mu'}^{(2)},
\label{eq:Umu0ma} \\
\tilde{V}_{0\mu',m\alpha}
&=&
\frac{1}{\mu_0} \left( \frac{\omega_{\rm p}^\prime}{c/n_{\rm bg}} \right)^2 B_{m\alpha,\mu'}^{(2)},
\label{eq:V0muma} \\
R_{m'\beta ,m\alpha}
&=&
\left( \frac{\omega_{\rm p}^\prime}{c/n_{\rm bg}} \right)^2 \int d^3\bm{k} \tilde{\varphi}_{m'} (-\bm{k})
\left( \frac{1}{- \bm{k}^2 + (n_{\rm bg} \omega/c)^2} \right) \tilde{\varphi}_m^\ast (\bm{k}) \delta_{\alpha \beta}
\nonumber \\
&=&
C_{m',m} \delta_{\alpha \beta}.
\label{eq:Rmbma}
\end{eqnarray}
Note that $\omega_{\rm p}^\prime = \omega_{\rm p}/n_{\rm bg}$ is the (bulk) plasma frequency in a material with the refractive index $n_{\rm bg}$.
The detail expression for $A_{\mu',\mu}^{(0,1,2)}$, $B_{m'\alpha,\mu}^{(1,2)}$, and $C_{m',m}$ are summarized in Appendix \ref{appx:NanoRod}.
}

{
\subsection{Exciation spectrum}
\label{sec:NRnumeric}
The matrix $\bar{\Xi} (\omega)$ of the self-consistent formulation in Eq.\ (\ref{eq:exSCmatXi}) describes
the spectrum for both the individual-like electron--hole excitations and the field-mediated collective-like excitation.
Then, we demonstrate the eigenvalues and determinant of the matrix $\bar{\Xi} (\omega)$ for
the nanorod as functions of the real component of frequency, $\omega = \omega_{\rm r} + i \omega_{\rm i}$.
By modulating the field-mediated couplings to reduce the T field contribution as shown in the following numerical demonstrations,
we find that the T field contributes significantly to form the collective-like excitation and
enlarges the radiative width related to the energy transfer between the excitations.
}

{
We consider $L_x = 10\, \mathrm{nm}$, $L_y = 15\, \mathrm{nm}$, and $L_z = 200\, \mathrm{nm}$ rectangular nanorod.
The Fermi energy of conduction electron is set at $\varepsilon_{\rm F} = 0.5\, \mathrm{eV}$.
The effective mass is $m^* = 0.02m_{\rm e}$ with $m_{\rm e}$ being the electron mass in vacuum.
Such an effective mass is obtained for InSb.
The background refractive index is taken as $n_{\rm bg} = 5$.
For a typical size scale $L_0 = 100\, \mathrm{nm}$, an order of the confinement energy is
$E_0 = \hbar^2 \pi^2 / (2 m^* L_0^2) \simeq 1.88\, \mathrm{meV}$.
Then, the electron--hole excitation energy is
$\hbar \omega_{\mu 0} = E_0 \left\{ (n_x^2 + n_y^2 + n_z^2) - (\bar{n}_x^2 + \bar{n}_y^2 + \bar{n}_z^2) \right\}$
with $\mu = ({\rm e}\mu ,{\rm h}\bar{\mu}) = (\{n_x,n_y,n_z\},\{\bar{n}_x,\bar{n}_y,\bar{n}_z\})$ satisfying
$E_0 (\bar{n}_x^2 + \bar{n}_y^2 + \bar{n}_z^2) \le \varepsilon_{\rm F} < E_0 (n_x^2 + n_y^2 + n_z^2)$.
We set $\gamma = 0.1 \times E_0$ as an infinitesimal value.
In bulk, the electron density is evaluated as $n_0 = (2m^* \varepsilon_{\rm F}/\hbar^2)^{3/2}/(3\pi^2)$.
Then, the bulk plasma frequency for above parameters is
$\hbar \omega_{\rm p}^\prime = \hbar \sqrt{e^2 n_0/(\varepsilon_{\rm bg} m^*)} \simeq 0.112\, \mathrm{eV}$.
}

\begin{figure}
\includegraphics[width=120mm]{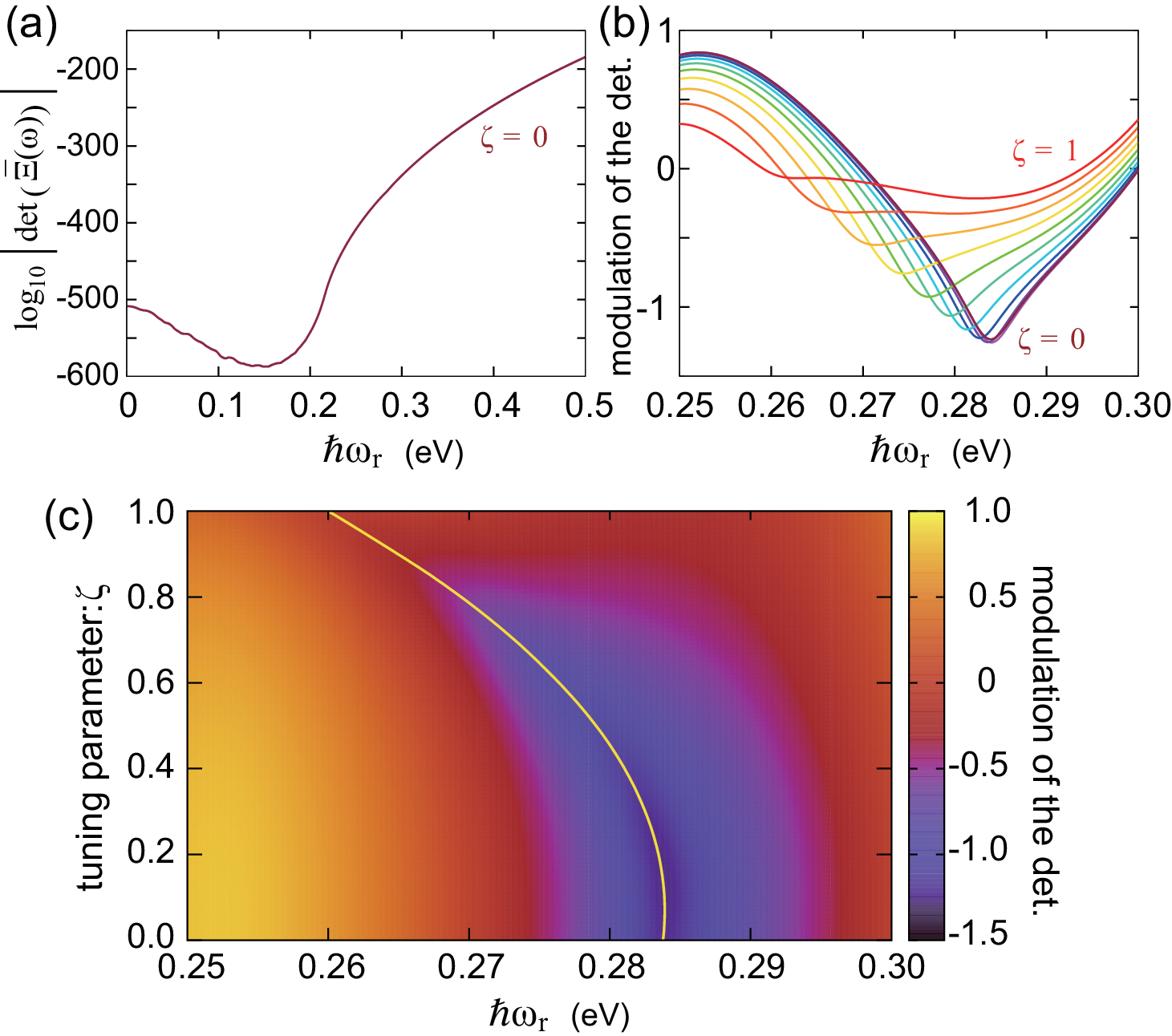}
\caption{
{
(a) Numerical results of ${\rm det} \left( \bar{\Xi} (\omega) \right)$ as a function of $\omega_{\rm r}$ with
$\omega = \omega_{\rm r} + i \omega_{\rm i}$ when the T field contribution is reduced ($\zeta = 0$).
A parameter $\zeta$ tunes the T components of the Green's function $\bar{\mathcal{G}}_{\bm{k}}$.
$\zeta = 1 (0)$ corresponds to a full (no transverse field) calculation.
The energy dimension of $\bar{\Xi} (\omega)$ is normalized.
The imaginary part of frequency is introduced to prevent mathematical divergence and set at $\hbar \omega_{\rm i} = 0.002\, \mathrm{eV}$.
(b) Modified plot of $\log \left| {\rm det} \left( \bar{\Xi} (\omega) \right) \right|$ in (a) to see clearly the collective excitation.
The determinant is subtracted by a harmonic function $f(x) \simeq -6580.9077 x^2 + 5034.8828 x - 1256.5965$ with
$x=\hbar \omega /(1\, \mathrm{eV})$, which is deduced from the values at $x = 0.95$, $1.05$, and $1.15$.
Lines indicate $\log \left| {\rm det} \left( \bar{\Xi} (\omega) \right) \right|$ from $\zeta = 1$ to $0$ by $0.1$.
(c) Color plot of modified $\log \left| {\rm det} \left( \bar{\Xi} (\omega) \right) \right|$ in (b).
A line indicates the position of $\omega_{\rm r}$ satisfying a real part of the eigenvalues of
$\bar{\Xi} (\omega)$ being zero, ${\rm Re} \left[ \xi_j (\omega) \right]=0$.
}
}
\label{fig:determinant}
\end{figure}

{
We focus on the excitation with $n_x = \bar{n}_x$, $n_y = \bar{n}_y$, and $n_z = \bar{n}_z +1$ to consider
the individual-like and collective-like excitation spectrum at a small wavenumber
$|\bm{q}| = |\bm{k}_{{\rm e} \mu} - \bm{k}_{{\rm h} \bar{\mu}}| = \pi /L_z$.
Because of stronger confinements in the $x$- and $y$-directions than that in the $z$-direction,
a subband structure characterized by index $n_{x,y}$ is formed; hence, our consideration corresponds to only the intra-subband excitation.
The excitation spectrum should be evaluated from the zero points of the eigenvalues of $\bar{\Xi} (\omega)$ in
the complex plane of $\omega = \omega_{\rm r} + i \omega_{\rm i}$.
In this demonstration, however, we fix the imaginary part of frequency at $\hbar \omega_{\rm i} = 0.002\, \mathrm{eV}$ for the visibility
and discuss the eigenvalues and determinant of $\bar{\Xi} (\omega)$ when $\omega_{\rm r}$ is swept.
First, we consider the determinant when the T field contribution is reduced from the matrix $\bar{\Xi} (\omega)$ to see clearly
the individual-like and collective-like excitations since the T field generally causes the radiative correction
and enlarges the imaginary components of the eigenvalues.
Then, we introduce a parameter $\zeta$ to tune the T component in the Green's function $\bar{\mathcal{G}}_{\bm{k}} (\omega)$,
i.e., we set $\tilde{K}_{\mu' 0,0\mu} = \zeta^2 A_{\mu',\mu}^{(2)} + \zeta A_{\mu',\mu}^{(1)} + A_{\mu',\mu}^{(0)}$,
where the order of $\zeta$ corresponds to that of current density in the matrix $\bar{\Xi} (\omega)$.
Namely, $\zeta = 1$ and $0$ signify a fully consideration and only the Coulomb interaction, respectively.
Figure \ref{fig:determinant}(a) represents $\log \left| {\rm det} \left( \bar{\Xi} (\omega) \right) \right|$ when $\zeta =0$.
The energy levels of the electron--hole excitations at $|\bm{q}| = \pi /L_z$ are distributed in $\hbar \omega_{\mu 0} < 0.22\, \mathrm{eV}$.
The eigenvalues of $\bar{\Xi} (\omega)$ indicates zero points at $\omega_{\rm r} \approx \omega_{\mu 0}$, which
suppresses strongly the determinant and shows jagged behavior at $\hbar \omega_{\rm r} < 0.22\, \mathrm{eV}$ in Fig.\ \ref{fig:determinant}(a).
In the present calculation, the imaginary frequency and the field-mediated couplings broaden the determinant dips.
Hence, it is difficult to find respective dip structures due to the respective electron--hole excitations.
}

{
Above the level distribution, the determinant increases monotonically with $\omega_{\rm r}$, where
the signature of a dip structure due to the collective excitation is difficult to be found from the determinant in this scale.
Then, we modulate the plot to see a signature of the collective-like excitation clearly by
subtracting a harmonic smooth function deduced from the plot in Fig.\ \ref{fig:determinant}(a).
Figures \ref{fig:determinant}(b) and (c) exhibit the modulated plot of $\log \left| {\rm det} \left( \bar{\Xi} (\omega) \right) \right|$,
which shows a shift of dip structure when the tuning parameter is tuned from $\zeta = 0$ to $1$.
By tuning $\zeta$, we can discuss a contribution of the T field to the construction of collective-like excitation.
Note that in our formulation, the existence of collective (plasmon) mode is not supposed in the Hamiltonian in an empirical way.
If the plasmon mode was assumed as an empirical model, the radiative T field would simply contribute to
the radiative shift of the assumed plasmon energy.
On the contrary, in our result, the collective(-like) mode appears in the deductive process from a cooperation of
the electron--hole excitations via both L and T fields.
Then, the obtained shift of collective-like excitation spectrum in Fig.\ \ref{fig:determinant}(b) contains
not only the radiative shift but also contributions by an additional mechanism to form the collective mode by the L and T fields.
If only the Coulomb interaction is considered ($\zeta = 0$), the modulated plot shows the dip structure due to
the collective-like excitation at $\hbar \omega_{\rm r} \approx 0.284\, \mathrm{eV}$ although
its depth is much smaller than the determinant in Fig.\ \ref{fig:determinant}(a).
Note that generally the plasmon excitation energy in the nanostructure is largely blue-shifted from $\omega_{\rm p}$ in bulk.
When $\zeta = 0$, the dip width is attributed to the imaginary part of the frequency.
If $\omega_{\rm i}$ is reduced, the dip structure becomes sharp and shows divergence at $\omega_{\rm i} = 0$ (not shown).
}

{
With a tuning of $\zeta$, in addition to the slight shift of dip position, the depth decreases and the width broadens.
It signifies that the T-field-mediated interaction between the excitations induces an energy dissipation.
When $\zeta = 1$, however, the dip structure is not distinguished in Fig.\ \ref{fig:determinant}(b);
hence, the energy of collective-like excitation cannot be evaluated from the determinant.
Then, the zero position of (real part of) the eigenvalue of $\bar{\Xi}$ corresponding to the collective-like excitation,
${\rm Re} \left[ \left( \bar{\Xi} (\omega) \right)_j \right] =0$, is examined in Fig.\ \ref{fig:determinant}(c).
The position also shifts with $\zeta$ from $\hbar \omega_{\rm r} \approx 0.284\, \mathrm{eV}$ to $\approx 0.260\, \mathrm{eV}$.
The collective-like excitation is well defined even at $\zeta = 1$ although the determinant does not indicate the distinct signature.
The shift of collective-like excitation with the increase of $\zeta$ is harmonic behavior.
Thus, the current--current interaction $A_{\mu',\mu}^{(2)}$ is dominant than the current--charge interaction $A_{\mu',\mu}^{(1)}$.
The contribution of the T component to the collective-like excitation is not negligible,
especially for the hybridization and the energy transfer between the individual-like and collective-like excitations.
A generation of the hot electron and hole should be described by the individual excitations.
Thus, the hybridization by the T field would be an important ingredient in the hot carrier generation from the plasmon excitation. 
}

\section{Discussion}
\label{sec:discussion}
In the self-consistent formulation in Eq.\ (\ref{eq:selfA}) and its matrix form (\ref{eq:exSCmat}) with Eq.\ (\ref{eq:exSCmatXi}),
the optical response of the nanostructures is described in terms of $\bm{\mathcal{J}}_{\nu \mu} (\bm{x})$.
The electron eigenstates $|\mu \rangle$ contribute to the optical response via $\bm{\mathcal{J}}_{\nu \mu} (\bm{x})$ in the susceptibility.
When one prepares the $|\mu \rangle$ of $\hat{H}_0$, including the Coulomb interaction at equilibrium,
the nanostructure optical responses, e.g., the plasmon--polariton and SPE spectra,
are obtained within a theoretical framework for the prepared eigenstates.
The first-principles calculation for a nanostructure or a nanoscale cluster of atoms was
developed~\cite{Kumar19,Zhang17,Chapkin18,Senanayake19,Ma15,Schira19,Lu19,Svendsen21},
where the electronic states with the electron--electron interaction (the L component of the EM field) are evaluated precisely.
In {many} previous studies, however, the T field was not self-consistently considered.
In our formulation, both components of the EM fields are determined self-consistently on the mesoscopic scale.
Further, if one employs the equilibrium states evaluated in the first-principles calculation in our formulation,
the T fields generated by the electronic responses and the electron--electron interactions are taken into account~\cite{com1}.
Hence, our approach provides an important and convenient framework for investigations of mesoscopic plasmonics and photonics.

The present formulation is based on the classical Maxwell's equations for the incident and induced EM fields.
As the induced L field describes part of the electron--electron interaction,
the quantum Maxwell's equations for the EM fields are required for the higher-order electron correlations.
For the linear response framework discussed in this paper, however, the classical treatment of the fields is equivalent to the quantum treatment.

Let us emphasize the difference in the four-vector representation of Eq.\ (\ref{eq:exSCmat}) compared to
the conventional nonlocal and self-consistent theory~\cite{book:cho1}.
In our formulation, both the vector and scalar potentials are fully considered.
The Coulomb gauge separates the L and T components in
{
$\bm{\mathcal{A}} = (\bm{A},-\phi /c)$,
}
which directly provides a picture of the L and T component mixing due to the nanostructure in the self-consistent relation in Eq.\ (\ref{eq:exSCmat}).
For the densities $\bm{\mathcal{J}} = (\bm{j}, c\rho)$ in the constitutive equation (\ref{eq:cons4vec}),
the current is induced not only by the T field, but also by the L field though
{
$\hat{H}^\prime \sim -\left( \hat{\bm{j}} \cdot \bm{A} - \hat{\rho} \phi \right)$.
}
The charge density is also attributed to both the L and T fields.
These characteristics are due to the off-diagonal elements in the $4 \times 4$ susceptibility $\bar{\mathcal{X}}$.
{In the Maxwell's equations (\ref{eq:formalA4vec})}, the L field $\phi$ is generated by the charge only,
whereas both the current and charge generate the T field $\bm{A}$.
This ``cross generation'' of LT components is essential physics for the nanostructure and is enlarged
when the nanostructure enhances its $\bm{\chi}_{j\rho}$ and $\bm{\chi}_{\rho j}$ due to the nonlocality.
The present formulation might provide a guideline for an enhancement of the plasmon resonance since
the generation of the current and charge densities is related to the collective-like and individual-like excitations.

{
We have demonstrated a single nanorod as an example of the applications in Sec.\ \ref{sec:NRnumeric}.
For rectangular nanorods, the charge and current densities in the radiative correlation matrices are analytically obtained.
Such analytical expressions present a clear understanding of the relation between the excitations and the induced densities.
The matrix components were evaluated numerically, which exhibits the practicality of our formulation.
The numerical results reveal the shift of collective-like excitation spectrum and enhanced radiative width due to the T field contribution.
In our formulation and model, the collective-like excitation appears in the deductive process from
the field-mediated interaction between the electron--hole excitations.
Then, our results suggest a new mechanism for the collective(-like) excitation.
}
{
In the present demonstration, we consider the individual excitations with only small wavenumber
$|\bm{q}| = |\bm{k}_{{\rm e} \mu} - \bm{k}_{{\rm h} \bar{\mu}}| = \pi /L_z$.
This corresponds to the intra-subband excitation for strong confinement in the $x$- and $y$-directions.
However, if the system is larger, or there are several interacting nanostructures in the $xy$-plane,
the charge and current deviations might be important.
In such situations, the inter-subband excitations becomes essential.
Moreover, the T field contributes to the coherent coupling between the collective-like and individual-like excitations.
In most previous studies that describe hot carrier generation caused by the plasmon excitation, 
the energy transfer is unidirectional~\cite{Zhang14,Ross15,Besteiro17,Manjavacas14,Chapkin18}.
However, if the coherent coupling between the collective and individual excitations is large,
a bidirectional energy transfer~\cite{Ma15,You18} can be effective in the nanostructures.
To discuss such coherent coupling mechanism, we will extend the model with large wavenumber for
the individual excitations considering the energy closing with the collective-like excitation in our future study.
}

{As a further discussion, based on our formulation,}
the optical responses of electrons are described in terms of the current and charge densities.
This representation is useful for considering the responses to the L and T field components.
In many previous studies, the optical properties were discussed in terms of the plasmons and SPEs.
In nanostructures featuring nonlocality, the plasmons and SPEs are coupled with each other.
Hence, the relation between the densities ($\bm{j},c \rho$) and the collective and individual excitations is complicated.
The excitation energy spectrum is evaluated from the eigenvalues of $\bar{\Xi} (\omega)$ in Eq.\ (\ref{eq:exSCmatXi})
as the excitations correspond to the poles of the system with the radiative field.
The densities $\bm{\mathcal{J}}$ and induced fields $\bm{\mathcal{A}}$ are described in terms of the excitations.
This is an advantage over another existing treatment of the nonlocal effect~\cite{Pitarke07}.
Moreover, from the poles, coherent coupling between the plasmon (collective-like) excitation
and the carrier generation due to the SPE can be investigated.

The coherent coupling of the collective-like and individual-like excitations in the nanostructures
gives rise to a bidirectional energy transfer~\cite{Ma15,You18}.
The plasmon and individual excitations form discrete and continuous spectra, respectively.
Therefore, their coupling may demonstrate the Fano resonance as evidence of coherent coupling.
In such a scenario, the self-consistency of the L and T total fields and the induced charge and current densities are important.
In addition to the energy, the electron wavefunction in the nanostructures may significantly affect the coupling.
Therefore, our microscopic-theory-based formulation reveals such light--nanostructure interaction and its enhancement.
As a multi-dimensional integral is included in Eqs.\ (\ref{eq:KJGJ})--(\ref{eq:exSFYA}), the presence of numerous electrons,
even in submicro-scale metallic structures, would make a feasibility of numerical calculations difficult.
The approach based on the first-principles calculation~\cite{Ma15,You18} has more limited applications than
the analytical approach based on our formulation.
Note also that we can overcome this difficulty for several nanostructures, e.g., a 2D sheet (including graphene) and a rectangular rod.

The development based on our microscopic approach are not limited to several adaptable nanostructures,
but can be applied to arbitrary nanostructures by considering appropriate approximations to restrict the bases.
Our approach can also be applied to an array of two or more nanostructure units,
in which the nonlocal effect is enlarged by the nanogap structure.
The collective(-like) excitations in the respective nanostructures are coupled with each other by the T and L fields.
Then, our formulation will contribute to reveal a coherent energy transfer between the nanostructures and its enhancement.

\section{Summary}
\label{sec:summary}
To correctly describe the coherent coupling between the collective plasmon excitations and the individual single-particle
excitations via a transverse electric field, which is an aspect that has been neglected in previous studies,
we developed a self-consistent and microscopic nonlocal formulation for the light--matter interaction in plasmonic nanostructures.
Our formulation is based on linear response theory with a nonlocal susceptibility and
the classical Maxwell's equations describing the transverse and longitudinal electromagnetic fields.
The nonlocal susceptibility $\bar{\mathcal{X}} (\bm{x},\bm{x}')$ is obtained from
the interaction Hamiltonian and the eigenstates of the nanostructure.
The Coulomb interaction is included not only in the non-perturbative Hamiltonian $\hat{H}_0$,
but also in the longitudinal electric fields {caused by the charge density related to the collective and individual excitations.}
In the formulation, the longitudinal and transverse components of the fields and optically induced responses are described in
terms of the four-vector potential $\bm{\mathcal{A}} (\bm{x},t)$ and density $\bm{\mathcal{J}} (\bm{x},t)$
with the Green's function $\bar{\mathcal{G}} (\bm{x},\bm{x}')$ in the four-vector space.
The self-consistent equation is rewritten in matrix form with the incident and induced fields, $Y^{(0,\pm)}_{\mu \nu}$ and $X^{(\pm)}_{\mu \nu}$.
From the poles of this formulated matrix $\bar{\Xi} (\omega)$ for the radiative correction, the excitation spectrum can be examined.
{
The developed formulation can be applied to various frameworks by
utilizing the electronic states obtained according to those respective frameworks.
We examine the excitation spectrum for a rectangular nanorod to demonstrate the numerical feasibility of our formulation.
The solutions of formulated matrix $\bar{\Xi} (\omega)$ provides the collective-like excitation spectrum, where
the transverse-field-mediated interaction contributes non-negligiblly to
the coherent coupling between the individual-like and collective-like excitations.
}

Our formulation could be combined with real-time density functional theory based on
the first-principles calculation, in which the electron--electron interaction is considered accurately.
The four-vector formulation can describe the effects of transverse fields on the eigenstates given by
the first-principles calculation at each step in a real-time simulation.
In this treatment, the exchange correlation and quantum fluctuation can be partially considered via the excited states.
Application of the present formulation to various types of metallic meso- and nanostructures will facilitate
theoretical understanding of the interplay between the collective(-like) and individual(-like) excitations in nanoscale systems,
which will aid the design of metallic systems to control photo-induced hot carrier generation.

\begin{acknowledgments}
T.Y. and H.I. are supported by JSPS KAKENHI JP16H06504 in Scientific Research on Innovative Areas,
``Nano-Material Optical-Manipulation.''
T.Y. is supported by JSPS KAKENHI 18K13484.
H.I. is supported by JSPS KAKENHI 18H01151.
\end{acknowledgments}

\appendix
{
\section{Another treatment of Hamiltonian}
\label{appx:anotherH}
In Sec.\ \ref{sec:Hamiltonian}, we separate the Hamiltonian into a static component including a part of
the Coulomb interaction, $\hat{H}_0$, and a light-induced component, $\hat{H}^\prime_{\rm ind}$.
In this description, the Coulomb interaction is included in the non-perturbative and perturbative Hamiltonian.
In this appendix, we consider another treatment for the Coulomb interaction and compare the two descriptions.
}

{
The total Hamiltonian $\hat{H}$ subtracting a constant value $U_0 = \int d\bm{x} \rho_0 \phi_{\rm ext}$
can be read as following two descriptions,
\begin{eqnarray}
\hat{H}
&=& \left( \hat{H}_0 + \hat{H}_{\rm p-p}^\prime \right)
 +  \left( \hat{H}_{\rm int} + \hat{H}_{\rm ext} \right)
= \hat{H}_0^\prime + \hat{H}_{\rm int}^\prime,
\label{eq:Hamiltonian2} \\
&=& \hat{H}_0
 +  \left( \hat{H}_{\rm int} + \hat{H}_{\rm ext} + \hat{H}_{\rm p-p}^\prime \right)
= \hat{H}_0 + \hat{H}_{\rm ind}^\prime.
\label{eq:Hamiltonian3}
\end{eqnarray}
Here, $\hat{H}_{\rm p-p}^\prime = \hat{H}_{\rm p-p} - U_0$.
}

{
In the former description of Eq.\ (\ref{eq:Hamiltonian2}),
the electron--electron interaction due to the L field caused by
the light-induced charge density is fully included in the non-perturbative Hamiltonian,
$\hat{H}_0^\prime \equiv \hat{H}_0 + \hat{H}_{\rm p-p}^\prime$.
The interaction with an external L field is in the perturbative one,
\begin{eqnarray}
\hat{H}_{\rm int}^\prime
&\equiv& \hat{H}_{\rm int} + \hat{H}_{\rm ext} \nonumber \\
&=& - \int d\bm{x} \left[
\hat{\bm{j}} (\bm{x},t) \cdot \bm{A} (\bm{x},t) - \hat{\rho} (\bm{x},t) \phi_{\rm ext} (\bm{x},t)
\right].
\label{eq:HTextL}
\end{eqnarray}
Here, both the incident and induced T fields are coupled with the charge current density.
Because the perturbative Hamiltonian is time-dependent via $\hat{H}_{\rm p-p}$ in Eq.\ (\ref{eq:Hpp2nd}),
one must prepare a quite complex and time-dependent basis $|\mu (t) \rangle$.
Thus, the evaluation of susceptibility by the time-dependent basis becomes complicated.
The induced Coulomb interaction contributes to the induced charge and current densities via the susceptibility.
This approach is reasonable if the self-consistent relation between the constitutive and Maxwell's equations is not considered.
However, for the self-consistent equation with this description,
the constitutive equation includes $\hat{\phi}_{\rm{mat}}$ and $\hat{\phi}_{\rm{pol}}$,
simultaneously, the interaction between the induced charges is also considered in the Maxwell's equations.
Then, a double count problem of the electron--electron interaction must also be handled.
Therefore, because of the basis complexity and the double count problem, this approach is not useful to
analyze the relation between the L and T components of the induced fields and densities.
}

{
For the latter description of Eq.\ (\ref{eq:Hamiltonian3}), however,
the non-perturbative Hamiltonian $\hat{H}_0$ and its basis $|\mu \rangle$ are static.
All induced effects by the light irradiation are included in the perturbative one,
\begin{equation}
\hat{H}_{\rm ind}^\prime = - \int d\bm{x} \left[
\hat{\bm{j}} (\bm{x},t) \cdot \bm{A} (\bm{x},t) - \delta \hat{\rho} (\bm{x},t)
\Big( \phi_{\rm ext} (\bm{x},t) + \phi_{\rm pol} (\bm{x},t) \Big)
\right].
\label{eq:HTfullL}
\end{equation}
Note that although $\phi_{\rm pol}$ is attributed to the Coulomb interaction between
the internal charges of the matter, it arises only when the external fields are applied.
Because of the electric neutrality of the matter, $\hat{\phi}_{\rm{mat}}$ is not incorporated in the Maxwell's equations.
Therefore, no treatment of the double counting of the electron--electron interaction is required
as the induced Coulomb interaction is irrelevant to the susceptibility in the constitutive equation
and considered only in the Maxwell's equations.
}

\begin{table}[t]
\begin{tabular}{c|c||c|c|c|c|c}
non-perturb.\ H & perturb.\ H & basis $| \mu \rangle$ & Coulomb int.\ & L field & T field & susceptibility \\
\hline \hline
$\hat{H}_0 + \hat{H}_{\rm{p-p}}^\prime$ & $\hat{H}_{\rm{int}}^\prime$ & time-dependent   & fully in $| \mu \rangle$     &
$\phi_{\rm{ext}}$                       & $\bm{A}$                    & time-dep.\ via $| \mu \rangle$ \\ \hline
$\hat{H}_0$                             & $\hat{H}_{\rm{ind}}^\prime$ & time-independent & partially in $| \mu \rangle$ &
$\phi_{\rm{pol}} + \phi_{\rm{ext}}$     & $\bm{A}$                    & static \\
\end{tabular}
\caption{
{
Treatment of Hamiltonian in Eqs.\ (\ref{eq:Hamiltonian2}) and (\ref{eq:Hamiltonian3}). Here,
$\phi_{\rm{ext}}$ describes an external (applied) L field, $\phi_{\rm{pol}}$ is an induced L field,
and $\bm{A}$ consists of the T field of both the applied and induced components.
}
}
\end{table}

{
The above discussion and classification are summarized in TABLE I and Figure 1.
}

\begin{figure}
\includegraphics[width=90mm]{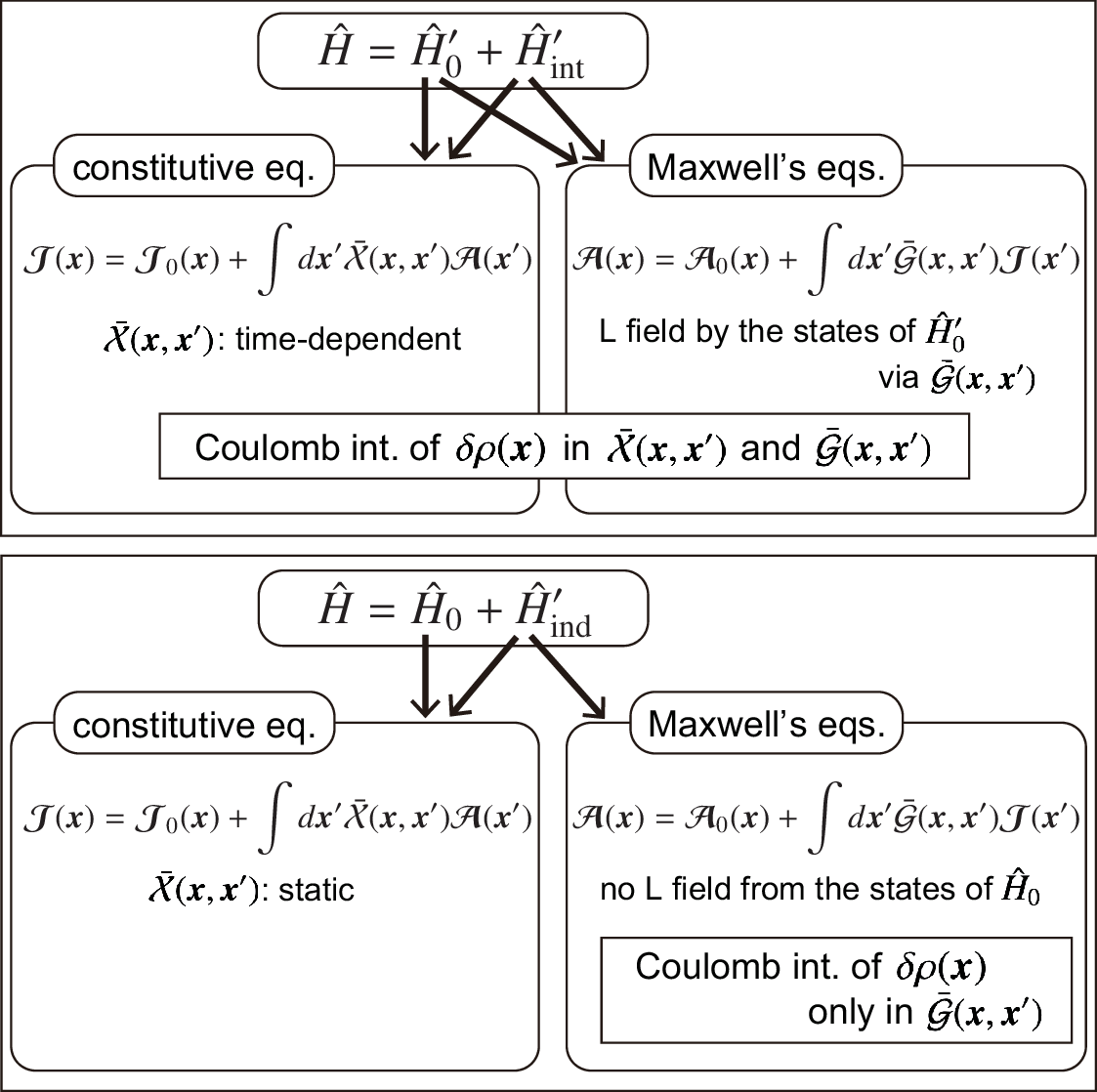}
\caption{
{
Schematic summary of two treatments of the Hamiltonian.
In the upper case, the electronic state of non-perturbative Hamiltonian has induced polarized charges.
The Coulomb interaction of the induced charges is included in both the constitutive and Maxwell's equations. 
Then, one has to take care of the double count of Coulomb interaction.
The lower case is our scheme, where the Coulomb interaction of the induced charges is
taken into account only in the Maxwell's equations.
Therefore, one does not need to consider the double count.
}
}
\label{fig:formulation}
\end{figure}

\section{L-T separation of Maxwell's equations}
\label{appx:LTMaxwell}
In the main text, we formulated the self-consistent equation in terms of
$\{ (\bm{A},-\phi /c),(\bm{j}, c \rho ) \}$ with the Coulomb gauge.
Let us consider another description in terms of
$\{ (\bm{A}^{\rm (T)},-\phi^{\rm (L)} /c) ,(\bm{j}^{\rm (T)},\bm{j}^{\rm (L)}) \}$.
Here, to emphasize the separation of the L and T components in the fields and densities,
we present the superscripts explicitly.
Equation (\ref{eq:CoulombJ}) can be expressed as 
\begin{eqnarray}
\left( \bm{\nabla}^2 - \frac{1}{c^2} \frac{\partial^2}{\partial t^2} \right) \bm{A}^{\rm (T)} (\bm{x},t)
&=& - \mu_0 \bm{j}^{\rm (T)} (\bm{x},t),
\label{eq:CoulombJT} \\
- \frac{1}{c^2} \frac{\partial}{\partial t} \bm{\nabla} \phi^{\rm (L)} (\bm{x},t)
&=& - \mu_0 \bm{j}^{\rm (L)} (\bm{x},t),
\label{eq:CoulombJL}
\end{eqnarray}
where $\bm{j} (\bm{x},t) = \bm{j}^{\rm (T)} (\bm{x},t) + \bm{j}^{\rm (L)} (\bm{x},t)$.
The scalar potential $\phi^{\rm (L)}$ generates the L component only.
Equation (\ref{eq:Coulombrho}) also describes the L only, where
\begin{equation}
\bm{\nabla}^2 \phi^{\rm (L)} (\bm{x},t) = - \frac{\rho^{\rm (L)} (\bm{x},t)}{\varepsilon_0}.
\label{eq:CoulombrhoL}
\end{equation}
Equations (\ref{eq:CoulombJL}) and (\ref{eq:CoulombrhoL}) satisfy the continuous relation for the L component,
\begin{equation}
\bm{\nabla} \cdot \bm{j}^{\rm (L)} (\bm{x},t) + \frac{\partial \rho^{\rm (L)} (\bm{x},t)}{\partial t} = 0.
\end{equation}
Let us refer to the T component in the scalar potential as $\phi^{\rm (T)}$.
The T component should be
$\bm{\nabla} \cdot \bm{E} = \bm{\nabla} \cdot \left( - \bm{\nabla} \phi^{\rm (T)} \right) = 0$.
Thus, it must have a linear dependence up to $\phi^{\rm (T)} (\bm{x},t) \sim a x + b$.
These terms cannot be induced from the internal charge density of the system.
We incorporate the external scalar potential $\phi_{\rm ext} (\bm{x},t)$ in
the interaction Hamiltonian (\ref{eq:Hindp}).
The T component in the external field, however, is included in the vector potential.
Therefore, the scalar potential induces the L component only, and we find
$\bm{E}^{\rm (L)} = - \bm{\nabla} \phi^{\rm (L)}$ and $\bm{E}^{\rm (T)} = - \partial_t \bm{A}^{\rm (T)}$ for
$\bm{E} = \bm{E}^{\rm (L)} + \bm{E}^{\rm (T)}$.

Such L and T field separation has been also discussed by Cho~\cite{book:cho2}, where
the scalar potential is reduced from the Hamiltonian (or Lagrangian) and
only the T components, $\bm{A}^{\rm (T)}$ and $\bm{j}^{\rm (T)}$, are included in the Maxwell's equations.
The L field, $\bm{E}^{\rm (L)}$, is treated as the external field only.

The representation of ($\bm{j}^{\rm (T)},\bm{j}^{\rm (L)}$) is a modification of ($\bm{j}, c\rho$).
It may be useful to separate the L and T components clearly in the source terms.
Moreover, the Green's function for Eqs.\ (\ref{eq:CoulombJT}) and (\ref{eq:CoulombJL}) is simpler than
Eq.\ (\ref{eq:solGreenMat}), as the off-diagonal components are absent.
However, for this representation, the constitutive equation (\ref{eq:cons4vec})
[and the perturbative Hamiltonian (\ref{eq:Hindp})] should be rewritten in terms of
($\bm{j}^{\rm (T)},\bm{j}^{\rm (L)}$).
Then, the susceptibility $\bar{\mathcal{X}}$ must be a $6 \times 4$ matrix,
which is not desirable in terms of the mathematics.

\section{Derivation of Green's function}
\label{appx:Green}
The Green's function for the Maxwell's equations is defined according to the differential operators in the equation,
which corresponds to the matrix $\bar{\mathcal{D}} (\bm{x} ;\omega)$ in Eq.\ (\ref{eq:Momega4vec}).
Therefore, the Green's function $\bar{\mathcal{G}} (\bm{x},\bm{x}^\prime ;\omega)$ is the $4 \times 4$ matrix,
\begin{equation}
\bar{\mathcal{D}} (\bm{x} ;\omega)
\bar{\mathcal{G}} (\bm{x},\bm{x}^\prime ;\omega)
= \bar{1} \delta (\bm{x} - \bm{x}^\prime).
\label{eq:Green4vec}
\end{equation}
Note that $\bar{\mathcal{D}} (\bm{x} ;\omega)$ is defined as Eq.\ (\ref{eq:D4vec}).
The formal solution of the Maxwell's equation is
\begin{equation}
\bm{\mathcal{A}} (\bm{x} ;\omega)
= \bm{\mathcal{A}}_0 (\bm{x} ;\omega)
- \mu_0 \int d\bm{x}^\prime \bar{\mathcal{G}} (\bm{x},\bm{x}^\prime ;\omega)
\bm{\mathcal{J}} (\bm{x}^\prime ;\omega).
\label{eq:formalA4vec2}
\end{equation}
The first term in Eq.\ (\ref{eq:formalA4vec2}) satisfies
\begin{equation}
\bar{\mathcal{D}} (\bm{x} ;\omega) \bm{\mathcal{A}}_0 (\bm{x} ;\omega) = 0.
\label{eq:A04vec2}
\end{equation}
To obtain an explicit form, $\bar{\mathcal{G}}$ is divided into the following matrix elements:
\begin{equation}
\bar{\mathcal{G}} (\bm{x},\bm{x}^\prime ;\omega)
=
\left( \begin{matrix}
\bar{g}_{AA} (\bm{x},\bm{x}^\prime ;\omega) &
\bm{g}_{A \phi} (\bm{x},\bm{x}^\prime ;\omega) \\
\bm{g}_{\phi A}^{\ {\rm t}} (\bm{x},\bm{x}^\prime ;\omega) &
g_{\phi \phi} (\bm{x},\bm{x}^\prime ;\omega)
\end{matrix} \right),
\label{eq:GreenMat4vec}
\end{equation}
where $\bar{g}_{AA}$ is a $3 \times 3$ matrix,
$\bar{g}_{\phi \phi}$ is scalar, and the other elements are vectors.

We have a mathematical preparation. The solution of
\begin{equation}
\left( \bm{\nabla}^2 + k^2 \right) g_k (\bm{x},\bm{x}^\prime) = \delta (\bm{x} - \bm{x}^\prime)
\label{eq:fkeq}
\end{equation}
is given as
\begin{equation}
g_k (\bm{x},\bm{x}^\prime) = - \frac{e^{ik|\bm{x} - \bm{x}^\prime|}}{4\pi |\bm{x} - \bm{x}^\prime|}.
\label{eq:fksol}
\end{equation}
Here, we avoid constant and linear terms, $\alpha_0 + \bm{\alpha}_1 \cdot \bm{x}$,
in Eq.\ (\ref{eq:fksol}), because of the boundary condition $g(|\bm{x}| \to \infty) = 0$.

{
The scalar potential component of the Green's function satisfies
\begin{equation}
- \bm{\nabla}^2 g_{\phi \phi} (\bm{x},\bm{x}^\prime ;\omega) = \delta (\bm{x} - \bm{x}^\prime),
\end{equation}
the solution of which is given by Eq.\ (\ref{eq:fksol}) as follows:
\begin{equation}
g_{\phi \phi} (\bm{x},\bm{x}^\prime ;\omega) = - g_0 (\bm{x},\bm{x}^\prime) = \frac{1}{4\pi |\bm{x} - \bm{x}^\prime|}.
\label{eq:solGreenphiphi}
\end{equation}
}
Further, $g_{\phi \phi}$ gives another matrix element $\bm{g}_{A \phi}$.
From Eq.\ (\ref{eq:Green4vec}),
\begin{equation}
\left( \bm{\nabla}^2 + k^2 \right) \bm{g}_{A \phi} (\bm{x},\bm{x}^\prime ;\omega)
- ik \bm{\nabla} g_{\phi \phi} (\bm{x},\bm{x}^\prime ;\omega) = 0.
\end{equation}
Here, $k=\omega/c$.
By substituting Eq.\ (\ref{eq:solGreenphiphi}),
\begin{equation}
\left( \bm{\nabla}^2 + k^2 \right) \bm{g}_{A \phi} (\bm{x},\bm{x}^\prime ;\omega)
= ik \bm{\nabla} g_{\phi \phi} (\bm{x},\bm{x}^\prime ;\omega)
= - \frac{ik}{4\pi} \frac{\bm{x} - \bm{x}^\prime}{|\bm{x} - \bm{x}^\prime|^3}.
\label{eq:GreenAphi4vec}
\end{equation}
This is an inhomogeneous equation, which is related to Eq.\ (\ref{eq:fkeq}).
The solution of $\bm{g}_{A \phi}$ is also obtained using Eq.\ (\ref{eq:fksol}), with
\begin{equation}
\bm{g}_{A \phi} (\bm{x},\bm{x}^\prime ;\omega)
= \bm{g}_k (\bm{x}) + \int d\bm{x}'' g_k (\bm{x},\bm{x}'')
\left( - \frac{ik}{4\pi} \frac{\bm{x}'' - \bm{x}^\prime}{|\bm{x}'' - \bm{x}^\prime|^3} \right).
\end{equation}
The first term satisfies $\left( \bm{\nabla}^2 + k^2 \right) \bm{g}_k (\bm{x}) = 0$.
It gives $\bm{g}_k (\bm{x}) = A_0 e^{i\bm{k} \cdot (\bm{x} - \bm{x}_0)}$ with $|\bm{k}| = \omega/c$.
Here, $A_0$ and $\bm{x}_0$ are constant. However, this term should be zero because of the boundary condition
at $|\bm{x}| \to \infty$.
Then, we obtain
\begin{equation}
\bm{g}_{A \phi} (\bm{x},\bm{x}^\prime ;\omega)
= \frac{ik}{(4\pi)^2} \int d\bm{x}''
\frac{e^{ik |\bm{x} - \bm{x}''|}}{|\bm{x} - \bm{x}''|}
\frac{\bm{x}'' - \bm{x}^\prime}{|\bm{x}'' - \bm{x}^\prime|^3}.
\label{eq:solGreenAphi}
\end{equation}
Next, we consider $\bar{g}_{AA}$ and $\bm{g}_{\phi A}^{\ {\rm t}}$.
From Eq.\ (\ref{eq:Green4vec}), $\bm{g}_{\phi A}^{\ {\rm t}}$ satisfies
\begin{equation}
\bm{\nabla}^2 \bm{g}_{\phi A}^{\ {\rm t}} (\bm{x},\bm{x}^\prime ;\omega) = 0.
\end{equation}
Note that $\bm{g}_{\phi A}^{\ {\rm t}}$ is a $1 \times 3$ row vector. The solution is 
$\bm{g}_{\phi A}^{\ {\rm t}} (\bm{x},\bm{x}^\prime ;\omega) = \left\{ \bar{\beta}_1 \bm{x} + \bm{\beta}_0 \right\}^{\rm t}$,
where $\bar{\beta}_1$ and $\bm{\beta}_0$ are a constant matrix and vector, respectively.
For the boundary condition, $\bm{\beta}_0 = 0$ and $\bar{\beta}_1 = 0$, and
\begin{equation}
\bm{g}_{\phi A}^{\ {\rm t}} (\bm{x},\bm{x}^\prime ;\omega) = 0.
\label{eq:solGreenphiA}
\end{equation}
The last element of the Green's function, $\bar{g}_{AA}$, satisfies
\begin{equation}
\left( \bm{\nabla}^2 + k^2 \right) \bar{g}_{AA} (\bm{x},\bm{x}^\prime ;\omega)
= \bar{1} \delta (\bm{x} - \bm{x}^\prime),
\end{equation}
which is equivalent to Eq.\ (\ref{eq:fkeq}), and the solution is
\begin{equation}
\bar{g}_{AA} (\bm{x},\bm{x}^\prime ;\omega)
= \bar{1} g_k (\bm{x},\bm{x}^\prime)
= - \frac{e^{ik|\bm{x} - \bm{x}^\prime|}}{4\pi |\bm{x} - \bm{x}^\prime|} \bar{1}.
\label{eq:solGreenAA}
\end{equation}
Finally, Eqs.\ (\ref{eq:solGreenphiphi}), (\ref{eq:solGreenphiA}), (\ref{eq:solGreenAphi}),
and (\ref{eq:solGreenAA}) give the Green's function in matrix form:
\begin{equation}
\bar{\mathcal{G}} (\bm{x},\bm{x}^\prime ;\omega)
=
- \frac{1}{4\pi}
\left( \begin{matrix}
\frac{ \ e^{ik|\bm{x} - \bm{x}^\prime|} }{|\bm{x} - \bm{x}^\prime|} \bar{1} \ &
\frac{-ik}{4\pi} \int d\bm{x}'' \frac{e^{ik |\bm{x} - \bm{x}''|}}{|\bm{x} - \bm{x}''|}
\frac{\bm{x}'' - \bm{x}^\prime}{|\bm{x}'' - \bm{x}^\prime|^3} \\
 0 &
{
- \frac{1}{|\bm{x} - \bm{x}^\prime|}
}
\end{matrix} \right).
\label{eq:solGreenMat-x}
\end{equation}
This Green's function provides a picture of the field (potential) generation due to
the electronic excitations (current and charge densities).

The Green's function in the $\bm{k}$-space representation (\ref{eq:solGreenMat_k}) is
obtained via a conventional Fourier transformation of Eq.\ (\ref{eq:Momega4vec}).
By applying
\begin{equation}
\bm{\mathcal{O}} (\bm{x} ;\omega)  = \frac{1}{(2\pi)^{\frac{3}{2}}} \int d\bm{k}
\tilde{\bm{\mathcal{O}}} (\bm{k} ;\omega) e^{i\bm{k} \cdot \bm{x}}
\end{equation}
for $\bm{\mathcal{O}} = \bm{\mathcal{A}}, \bm{\mathcal{J}}$, we find
\begin{eqnarray}
\bar{\mathcal{D}} (\bm{x} ;\omega) \left\{ \frac{1}{(2\pi)^{\frac{3}{2}}} \int d\bm{k}
\tilde{\bm{\mathcal{A}}} (\bm{k} ;\omega) e^{i\bm{k} \cdot \bm{x}} \right\}
&=&
\frac{1}{(2\pi)^{\frac{3}{2}}} \int d\bm{k}
\left( \begin{matrix}
\left( -\bm{k}^2 + \frac{\omega^2}{c^2} \right) \bar{1} &
{
\frac{\omega}{c} \bm{k}
} \\
0 &
{
\bm{k}^2
}
\end{matrix} \right)
\tilde{\bm{\mathcal{A}}} (\bm{k} ;\omega) e^{i\bm{k} \cdot \bm{x}}
\nonumber \\
&=&
\frac{1}{(2\pi)^{\frac{3}{2}}} \int d\bm{k} (-\mu_0)
\tilde{\bm{\mathcal{J}}} (\bm{k} ;\omega) e^{i\bm{k} \cdot \bm{x}}.
\end{eqnarray}
Hence, the Green's function is
\begin{equation}
\bar{\mathcal{G}}_{\bm{k}} (\omega) = \left( \begin{matrix}
\left( -\bm{k}^2 + \frac{\omega^2}{c^2} \right) \bar{1} &
{
\frac{\omega}{c} \bm{k}
}
\\
0 &
{
\bm{k}^2
}
\end{matrix} \right)^{-1}
=
\left( \begin{matrix}
\frac{1}{-\bm{k}^2 + \omega^2/c^2} \bar{1} &
- \frac{1}{\bm{k}^2} \frac{1}{-\bm{k}^2 + \omega^2/c^2} \frac{\omega}{c} \bm{k} \\
0 &
{
\frac{1}{\bm{k}^2}
}
\end{matrix} \right).
\end{equation}

The formal solutions of $\bm{\mathcal{A}} = (\bm{A},-\phi /c)$ can also be obtained from
the separated Maxwell's equations in Eqs.\ (\ref{eq:CoulombJ}) and (\ref{eq:Coulombrho}).
The latter equation gives
\begin{equation}
\phi (\bm{x} ;\omega) = \phi_0 (\bm{x} ;\omega)
+ \frac{1}{\varepsilon_0} \int d\bm{x}^\prime g_{\phi} (\bm{x},\bm{x}^\prime ) \rho (\bm{x}^\prime ;\omega).
\label{eq:formalphi}
\end{equation}
The scalar Green's function $g_{\phi} (\bm{x},\bm{x}^\prime )$ is equivalent to
$g_{\phi \phi} (\bm{x},\bm{x}^\prime )$ in Eq.\ (\ref{eq:solGreenphiphi}).
The first term satisfies $\bm{\nabla}^2 \phi_0 (\bm{x};\omega) =0$.
In the Coulomb gauge, however, the T field is described by $\bm{A} (\bm{x};\omega)$ only and, hence,
$\phi_0 =0$.
The solution of Eq.\ (\ref{eq:CoulombJ}) is
\begin{equation}
\bm{A} (\bm{x} ;\omega) = \bm{A}_0 (\bm{x} ;\omega)
- \mu_0 \int d\bm{x}^\prime \bar{G}_{\bm{A}} (\bm{x},\bm{x}^\prime ;\omega) \bm{j} (\bm{x}^\prime ;\omega)
\label{eq:formalA}
\end {equation}
with $\left( \bm{\nabla}^2 + \frac{\omega^2}{c^2} \right) \bm{A}_0 (\bm{x} ;\omega) = 0$.
Here, the dyadic Green's function $\bar{G}_{\bm{A}}$ for the vector potential is expressed as
\begin{eqnarray}
\bar{G}_{\bm{A}} (\bm{x},\bm{x}^\prime ;\omega)
&=& g_{\bm{A}} (\bm{x},\bm{x}^\prime ;\omega) \bar{1}
+ \int d\bm{x}'' g_{\bm{A}} ( \bm{x},\bm{x}'' ;\omega)
\bm{\nabla}_{\bm{x}''} g_\phi (\bm{x}'' ,\bm{x}^\prime) \bm{\nabla}_{\bm{x}^\prime} \cdot
\nonumber \\
&=& g_{\bm{A}} (\bm{x},\bm{x}^\prime ;\omega) \bar{1}
+ \int d\bm{x}'' \left[ \bm{\nabla}_{\bm{x}''} g_{\bm{A}} ( \bm{x},\bm{x}'' ;\omega) \right]
\left[ \bm{\nabla}_{\bm{x}^\prime} g_\phi (\bm{x}'' ,\bm{x}^\prime) \right] \cdot
\label{eq:GreenAscho2}
\end {eqnarray}
with
\begin{equation}
g_{\bm{A}} (\bm{x},\bm{x}^\prime ;\omega)
= - \frac{e^{i\frac{\omega}{c} |\bm{x} - \bm{x}^\prime|}}{4\pi |\bm{x} - \bm{x}^\prime|}.
\label{eq:solGreenA}
\end{equation}
For the derivation, Eq.\ (\ref{eq:formalphi}) is substituted and the continuous equation
$\bm{\nabla} \cdot \bm{j} (\bm{x};\omega) -i\omega \rho (\bm{x};\omega) = 0$ is used, as
the source term is described by $\bm{j} (\bm{x} ;\omega)$ only in this representation.
In the dyadic Green's function $\bar{G}_{\bm{A}} (\bm{x},\bm{x}^\prime ;\omega)$ in Eq.\ (\ref{eq:GreenAscho2}),
$g_\phi$ is included, which implies that the L component is related to the vector potential.
This seems strange at first glance, as the vector potential in Eq.\ (\ref{eq:formalA}) must satisfy $\bm{\nabla}_x \cdot \bm{A} = 0$.

To see it, we separate the T and L components of the current density,
$\bm{j} = \bm{j}^{\rm (T)} + \bm{j}^{\rm (L)}$.
This is a description in terms of
$\{ (\bm{A}^{\rm (T)},-\phi^{\rm (L)}/c),(\bm{j}^{\rm (T)},\bm{j}^{\rm (L)}) \}$,
which is discussed in Appendix \ref{appx:LTMaxwell}.
The formal solution for Eq.\ (\ref{eq:CoulombJT}) is given by $g_{\bm{A}} (\bm{x},\bm{x}^\prime ;\omega)$,
\begin{equation}
\bm{A}^{\rm (T)} (\bm{x} ;\omega) = \bm{A}_0^{\rm (T)} (\bm{x} ;\omega)
- \mu_0 \int d\bm{x}^\prime g_{\bm{A}} (\bm{x},\bm{x}^\prime ;\omega) \bm{j}^{\rm (T)} (\bm{x}^\prime ;\omega).
\label{eq:formalALT}
\end{equation}
Note that Eq.\ (\ref{eq:formalALT}) differs slightly from Eq.\ (\ref{eq:formalA}) in its second term on the r.h.s.
For the scalar potential, we apply the continuous relation, $i\omega \rho = \bm{\nabla} \cdot \bm{j}^{\rm (L)}$, to
Eq.\ (\ref{eq:formalphi}),
\begin{eqnarray}
\phi^{\rm (L)} (\bm{x} ;\omega)
&=& \frac{1}{i\omega \varepsilon_0} \int d\bm{x}^\prime g_\phi (\bm{x},\bm{x}^\prime )
\bm{\nabla}_{x^\prime} \cdot \bm{j}^{\rm (L)} (\bm{x}^\prime ;\omega).
\label{eq:formalphiLT}
\end{eqnarray}
The set of Eqs.\ (\ref{eq:formalALT}) and (\ref{eq:formalphiLT}) give the solutions of the Maxwell's equation for the L--T separation.
For the source term in Eq.\ (\ref{eq:formalALT}), we apply Eqs.\ (\ref{eq:CoulombJL}), (\ref{eq:formalphiLT}),
and the continuous relation,
\begin{eqnarray}
\bm{j}^{\rm (T)} (\bm{x}^\prime ;\omega)
&=&
\bm{j} (\bm{x}^\prime ;\omega) - \bm{j}^{\rm (L)} (\bm{x}^\prime ;\omega) \nonumber \\
&=&
\bm{j} (\bm{x}^\prime ;\omega) - \frac{-i\omega}{\mu_0 c^2} \bm{\nabla}_{x^\prime} \phi^{\rm (L)} (\bm{x}^\prime ;\omega) \nonumber \\
&=&
\bm{j} (\bm{x}^\prime ;\omega) + \bm{\nabla}_{x^\prime}
\int d\bm{x}'' g_\phi (\bm{x}^\prime,\bm{x}'') \bm{\nabla}_{x''} \cdot \bm{j}^{\rm (L)} (\bm{x}'' ;\omega) \nonumber \\
&=&
\int d\bm{x}'' \left\{ \bar{1} \delta (\bm{x}^\prime - \bm{x}'')
+ \bm{\nabla}_{x^\prime} g_\phi (\bm{x}^\prime,\bm{x}'') \bm{\nabla}_{x''} \cdot \right\} \bm{j} (\bm{x}'' ;\omega).
\end {eqnarray}
Then, Eqs.\ (\ref{eq:formalALT}) and (\ref{eq:formalA}) are equivalent.
Note that the last term can be rewritten as a dyadic function,
\begin{equation}
\bm{\nabla}_{x^\prime} g_\phi (\bm{x}^\prime,\bm{x}'') \bm{\nabla}_{x''} \cdot
= \overleftrightarrow{ \left( \bm{\nabla}_{x^\prime} \bm{\nabla}_{x^\prime} \cdot \right) }
g_\phi (\bm{x}^\prime,\bm{x}'').
\end{equation}
As $g_{\bm{A}} (\bm{x},\bm{x}^\prime ;\omega)$ depends only on $|\bm{x}-\bm{x}^\prime|$, we find
\begin{eqnarray*}
\bm{\nabla}_x \cdot \int d\bm{x}^\prime g_{\bm{A}} (\bm{x},\bm{x}^\prime ;\omega)
\bm{j}^{\rm (T)} (\bm{x}^\prime ;\omega),
&=&
- \int d\bm{x}^\prime \left( \bm{\nabla}_{x^\prime} \cdot g_{\bm{A}} (\bm{x},\bm{x}^\prime ;\omega) \right)
\bm{j}^{\rm (T)} (\bm{x}^\prime ;\omega), \\
&=&
\int d\bm{x}^\prime g_{\bm{A}} (\bm{x},\bm{x}^\prime ;\omega)
\left( \bm{\nabla}_{x^\prime} \cdot \bm{j}^{\rm (T)} (\bm{x}^\prime ;\omega) \right) \\
&=& 0.
\end {eqnarray*}
This follows the transverse vector potential, $\bm{\nabla}_x \cdot \bm{A} = 0$.
Therefore, the solutions (\ref{eq:formalphi}) and (\ref{eq:formalA}) obtained from the separated Maxwell's equations have no inquiries.

These treatments of the source term $\bm{j}$ and
the L and T field components complicate the theoretical framework.
In our four-vector representation, such complexity becomes clear as
the two components $(\bm{A} ,-\phi /c)$ of the fields and
the two source components $(\bm{j}, c\rho)$ are related to each other in matrix form.

{
\section{Derivation of the matrix form of self-consistent equation}
\label{appx:SCmatrix}
In this appendix, we describe the detail derivation and formulation of the matrix form of
self-consistent equation from Eq.\ (\ref{eq:selfA}),
\begin{eqnarray}
\bm{\mathcal{A}} (\bm{x};\omega)
&=&
\bm{\mathcal{A}}_0 (\bm{x};\omega)
- \mu_0 \int d\bm{x}' \bar{\mathcal{G}}(\bm{x},\bm{x}';\omega) \bm{\mathcal{J}}_0 (\bm{x}';\omega)
\nonumber \\
& & \hspace{20mm}
- \mu_0 \int d\bm{x}' \int d\bm{x}'' \bar{\mathcal{G}} (\bm{x},\bm{x}';\omega)
\bar{\mathcal{X}} (\bm{x}',\bm{x}'';\omega) \bm{\mathcal{A}} (\bm{x}'';\omega).
\end{eqnarray}
For the third term in the r.h.s., one substitutes the nonlocal susceptibility
$\bar{\mathcal{X}} (\bm{x}',\bm{x}'';\omega)$ in Eq.\ (\ref{eq:nonlocalsus4vec}),
and for the second term, one uses Eq.\ (\ref{eq:exvectorterm}).
By multiplying with $\left( \bm{\mathcal{J}}_{\nu' \mu'} (\bm{x}) \right)^{\rm t}$,
$\left( \bm{\mathcal{J}}_{\mu' \nu'} (\bm{x}) \right)^{\rm t}$, and
$\left( \varphi_i (\bm{x}) \bm{e}_\beta^{\rm t} \right)$ from the left
and integrating by $\bm{x}$, one obtains
\begin{eqnarray}
\left(\hbar \omega_{\nu' \mu'} - \hbar \omega - i\gamma \right) X_{\nu' \mu'}^{(-)}
&=&
Y_{\nu' \mu'}^{(0)}
+ \sum_{j,\alpha}       U_{\nu' \mu',j \alpha} X_{j\alpha}^{(A)}
- \sum_{\mu,\nu} \left[ K_{\nu' \mu',\mu \nu}  X_{\nu \mu}^{(-)}
                      + L_{\nu' \mu',\nu \mu}  X_{\mu \nu}^{(+)} \right],
\label{eq:exSCmatrix1} \\
\left(\hbar \omega_{\nu' \mu'} + \hbar \omega + i\gamma \right) X_{\mu' \nu'}^{(+)}
&=&
Y_{\mu' \nu'}^{(0)}
+ \sum_{j,\alpha}       V_{\mu' \nu',j \alpha} X_{j\alpha}^{(A)}
- \sum_{\mu,\nu} \left[ M_{\mu' \nu',\mu \nu}  X_{\nu \mu}^{(-)}
                      + N_{\mu' \nu',\nu \mu}  X_{\mu \nu}^{(+)} \right],
\label{eq:exSCmatrix2} \\
X_{i \beta}^{(A)}
&=&
Y_{i \beta}^{(A)}
+ \sum_{j,\alpha}       R_{i \beta,j \alpha} X_{j \alpha}^{(A)}
- \sum_{\mu,\nu} \left[ S_{i \beta,\mu \nu}  X_{\nu \mu}^{(-)}
                      + T_{i \beta,\nu \mu}  X_{\mu \nu}^{(+)} \right].
\label{eq:exSCmatrix3}
\end{eqnarray}
Here, the factors are defined in Eqs.\ (\ref{eq:KJGJ})--(\ref{eq:exSFYA}) in Sec.\ \ref{sec:x-rep}.
Note again that the excited states $\mu$ indicating an electron--hole pair include
their spin degrees of freedom. $\alpha, \beta$ takes only $x,y,z$ not $\phi$.
The factors, $K_{\nu' \mu' ,\mu \nu}$, $L_{\nu' \mu' ,\nu \mu}$,
$M_{\mu' \nu' ,\mu \nu}$, and $N_{\mu' \nu' ,\nu \mu}$ are mathematically related.
The matrix form of Eqs.\ (\ref{eq:exSCmatrix1})-(\ref{eq:exSCmatrix3}) is written as
\begin{eqnarray}
\left[ \left( \hbar \bar{\Omega} - (\hbar \omega + i\gamma) \bar{1} \right) + \bar{K} \right] \bm{X}^{(-)}
+ \bar{L} \bm{X}^{(+)}
&=&
\bm{Y}^{(0,-)} + \bar{U} \bm{X}^{(A)},
\label{eq:exSCmatfrom1} \\
\left[ \left( \hbar \bar{\Omega} + (\hbar \omega + i\gamma) \bar{1} \right) + \bar{N} \right] \bm{X}^{(+)}
+ \bar{M} \bm{X}^{(-)}
&=&
\bm{Y}^{(0,+)} + \bar{V} \bm{X}^{(A)},
\label{eq:exSCmatfrom2} \\
\left[ \bar{1} - \bar{R} \right] \bm{X}^{(A)}
&=&
\bm{Y}^{(A)} - \left[ \bar{S} \bm{X}^{(-)} + \bar{T} \bm{X}^{(+)} \right].
\label{eq:exSCmatfrom3}
\end{eqnarray}
$\bar{K}$ and $\bar{N}$ indicate the resonant and anti-resonant radiative correlations, respectively,
and $\bar{L}$ and $\bar{M}$ describes their couplings.
Here, one introduces an integer $N$ being a cut-off number of the states,
and defines the vectors of $X_{\nu \mu}^{(\pm)}$ for the fields as
\begin{equation}
\bm{X}^{(\pm)}
= \left( \begin{matrix}
\bm{X}^{(\pm)}_0 \\
\bm{X}^{(\pm)}_1 \\
\vdots \\
\bm{X}^{(\pm)}_N
\end{matrix} \right)
\ \ \ {\rm with} \ \ \
\bm{X}^{(-)}_\mu
= \left( \begin{matrix}
X^{(-)}_{0 \mu} \\
X^{(-)}_{1 \mu} \\
\vdots \\
X^{(-)}_{N \mu}
\end{matrix} \right)
\ \ , \ \
\bm{X}^{(+)}_\mu
= \left( \begin{matrix}
X^{(+)}_{\mu 0} \\
X^{(+)}_{\mu 1} \\
\vdots \\
X^{(+)}_{\mu N}
\end{matrix} \right).
\label{eq:Xpmvec}
\end{equation}
For the incident field $Y^{(0)}_{\nu \mu}$,
\begin{equation}
\bm{Y}^{(0,\pm)}
= \left( \begin{matrix}
\bm{Y}^{(0,\pm)}_0 \\
\bm{Y}^{(0,\pm)}_1 \\
\vdots \\
\bm{Y}^{(0,\pm)}_N
\end{matrix} \right)
\ \ \ {\rm with} \ \ \
\bm{Y}^{(0,-)}_\mu
= \left( \begin{matrix}
Y^{(0)}_{0 \mu} \\
Y^{(0)}_{1 \mu} \\
\vdots \\
Y^{(0)}_{N \mu}
\end{matrix} \right)
\ \ , \ \
\bm{Y}^{(0,+)}_\mu
= \left( \begin{matrix}
Y^{(0)}_{\mu 0} \\
Y^{(0)}_{\mu 1} \\
\vdots \\
Y^{(0)}_{\mu N}
\end{matrix} \right).
\label{eq:Ypmvec}
\end{equation}
Here, $\bm{Y}^{(0,-)}$ and $\bm{Y}^{(0,+)}$ correspond to each other by a permutation of the elements.
For the vector potential term in the current $X_{j\alpha}^{(A)}$,
\begin{equation}
\bm{X}^{(A)}
= \left( \begin{matrix}
\bm{X}^{(A)}_0 \\
\bm{X}^{(A)}_1 \\
\vdots \\
\bm{X}^{(A)}_M
\end{matrix} \right)
\ \ \ {\rm with} \ \ \
\bm{X}^{(A)}_j
= \left( \begin{matrix}
X^{(A)}_{jx} \\
X^{(A)}_{jy} \\
X^{(A)}_{jz}
\end{matrix} \right).
\label{eq:XAvec}
\end{equation}
For the incident field $Y_{j\alpha}^{(A)}$ ,
\begin{equation}
\bm{Y}^{(A)}
= \left( \begin{matrix}
\bm{Y}^{(A)}_0 \\
\bm{Y}^{(A)}_1 \\
\vdots \\
\bm{Y}^{(A)}_M
\end{matrix} \right)
\ \ \ {\rm with} \ \ \
\bm{Y}^{(A)}_j
= \left( \begin{matrix}
Y^{(A)}_{jx} \\
Y^{(A)}_{jy} \\
Y^{(A)}_{jz}
\end{matrix} \right).
\label{eq:YAvec}
\end{equation}
}

{
Next, the matrices in Eqs.\ (\ref{eq:exSCmatfrom1})-(\ref{eq:exSCmatfrom3}) are defined as follows.
For the energy differences $\omega_{\nu \mu}$,
\begin{equation}
\bar{\Omega}
= \left( \begin{matrix}
\bar{\Omega}_0 &                &        & \\
               & \bar{\Omega}_1 &        & \\
               &                & \ddots & \\
               &                &        & \bar{\Omega}_N
\end{matrix} \right)
\label{eq:Omega}
\end{equation}
is a diagonal matrix with
$\bar{\Omega}_\mu = {\rm diag} (\omega_{0\mu},\omega_{1\mu},\omega_{2\mu}\cdots,\omega_{N\mu})$.
For the matrices $\bar{K}$, $\bar{L}$, $\bar{M}$, and $\bar{N}$,
\begin{equation}
\bar{Q}
= \left( \begin{matrix}
\bar{Q}_{00} & \bar{Q}_{01} & \cdots & \bar{Q}_{0N} \\
\bar{Q}_{10} & \bar{Q}_{11} &        & \\
\vdots       &              & \ddots & \\
\bar{Q}_{N0} &              &        & \bar{Q}_{NN}
\end{matrix} \right)
\ \ (Q=K,L,M,N)
\end{equation}
with different subscript rules for $(N+1) \times (N+1)$ block matrices
\begin{eqnarray}
\bar{K}_{\mu' \mu}
&=& \left( \begin{matrix}
K_{0 \mu' ,\mu 0} & K_{0 \mu' ,\mu 1} & \cdots & K_{0 \mu' ,\mu N} \\
K_{1 \mu' ,\mu 0} & K_{1 \mu' ,\mu 1} &        & \\
\vdots            &                   & \ddots & \\
K_{N \mu' ,\mu 0} &                   &        & K_{N \mu' ,\mu N}
\end{matrix} \right),
\label{eq:Kmatrix} \\
\bar{L}_{\mu' \mu}
&=& \left( \begin{matrix}
L_{0 \mu' ,0 \mu} & L_{0 \mu' ,1 \mu} & \cdots & L_{0 \mu' ,N \mu} \\
L_{1 \mu' ,0 \mu} & L_{1 \mu' ,1 \mu} &        & \\
\vdots            &                   & \ddots & \\
L_{N \mu' ,0 \mu} &                   &        & L_{N \mu' ,N \mu}
\end{matrix} \right),
\label{eq:Lmatrix} \\
\bar{M}_{\mu' \mu}
&=& \left( \begin{matrix}
M_{\mu' 0,\mu 0} & M_{\mu' 0,\mu 1} & \cdots & M_{\mu' 0,\mu N} \\
M_{\mu' 1,\mu 0} & M_{\mu' 1,\mu 1} &        & \\
\vdots           &                  & \ddots & \\
M_{\mu' N,\mu 0} &                  &        & M_{\mu' N,\mu N}
\end{matrix} \right),
\label{eq:Mmatrix}
\end{eqnarray}
\begin{eqnarray}
\bar{N}_{\mu' \mu}
&=& \left( \begin{matrix}
N_{\mu' 0,0 \mu} & N_{\mu' 0,1 \mu} & \cdots & N_{\mu' 0,N \mu} \\
N_{\mu' 1,0 \mu} & N_{\mu' 1,1 \mu} &        & \\
\vdots           &                  & \ddots & \\
N_{\mu' N,0 \mu} &                  &        & N_{\mu' N,N \mu}
\end{matrix} \right).
\label{eq:Nmatrix}
\end{eqnarray}
For the matrix $\bar{R}$ related to the vector potential term in the current,
\begin{equation}
\bar{R}
= \left( \begin{matrix}
\bar{R}_{11} & \bar{R}_{12} & \cdots & \bar{R}_{1M} \\
\bar{R}_{21} & \bar{R}_{22} &        & \\
\vdots       &              & \ddots & \\
\bar{R}_{M1} &              &        & \bar{R}_{MM}
\end{matrix} \right)
\end{equation}
with $3 \times 3$ block matrices
\begin{equation}
\bar{R}_{ij}
= \left( \begin{matrix}
R_{ix,jx} & R_{ix,jy} & R_{ix,jz} \\
R_{iy,jx} & R_{iy,jy} & R_{iy,jz} \\
R_{iz,jx} & R_{iz,jy} & R_{iz,jz}
\end{matrix} \right).
\label{eq:Rmatrix}
\end{equation}
For the off-diagonal matrices $\bar{U}$ and $\bar{V}$,
\begin{equation}
\bar{Q}
= \left( \begin{matrix}
\bar{Q}_{01} & \bar{Q}_{02} & \cdots & \bar{Q}_{0M} \\
\bar{Q}_{11} & \bar{Q}_{12} &        & \\
\vdots       &              & \ddots & \\
\bar{Q}_{N1} &              &        & \bar{Q}_{NM}
\end{matrix} \right)
\ \ (Q=U,V)
\end{equation}
with $(N+1) \times 3$ block matrices
\begin{eqnarray}
\bar{U}_{\mu'j}
&=& \left( \begin{matrix}
U_{0\mu',jx} & U_{0\mu',jy} & U_{0\mu',jz} \\
U_{1\mu',jx} & U_{1\mu',jy} & U_{1\mu',jz} \\
             & \vdots       &              \\
U_{N\mu',jx} & U_{N\mu',jy} & U_{N\mu',jz}
\end{matrix} \right),
\label{eq:Umatrix} \\
\bar{V}_{\mu'j}
&=& \left( \begin{matrix}
V_{\mu'0,jx} & V_{\mu'0,jy} & V_{\mu'0,jz} \\
V_{\mu'1,jx} & V_{\mu'1,jy} & V_{\mu'1,jz} \\
             & \vdots       &              \\
V_{\mu'N,jx} & V_{\mu'N,jy} & V_{\mu'N,jz}
\end{matrix} \right)
\label{eq:Vmatrix}
\end{eqnarray}
and for $\bar{S}$ and $\bar{T}$,
\begin{equation}
\bar{Q}
= \left( \begin{matrix}
\bar{Q}_{10} & \bar{Q}_{11} & \cdots & \bar{Q}_{1N} \\
\bar{Q}_{20} & \bar{Q}_{21} &        & \\
\vdots       &              & \ddots & \\
\bar{Q}_{M0} &              &        & \bar{Q}_{MN}
\end{matrix} \right)
\ \ (Q=S,T)
\end{equation}
with $3 \times (N+1)$ block matrices
\begin{eqnarray}
\bar{S}_{i\mu}
&=& \left( \begin{matrix}
S_{ix,\mu 0}     & S_{ix,\mu 1}     &        & S_{ix,\mu N} \\
S_{iy,\mu 0}     & S_{iy,\mu 1}     & \cdots & S_{iy,\mu N} \\
S_{iz,\mu 0}     & S_{iz,\mu 1}     &        & S_{iz,\mu N}
\end{matrix} \right),
\label{eq:Smatrix} \\
\bar{T}_{i\mu}
&=& \left( \begin{matrix}
T_{ix,0\mu}     & T_{ix,1\mu}     &        & T_{ix,N\mu} \\
T_{iy,0\mu}     & T_{iy,1\mu}     & \cdots & T_{iy,N\mu} \\
T_{iz,0\mu}     & T_{iz,1\mu}     &        & T_{iz,N\mu}
\end{matrix} \right).
\label{eq:Tmatrix}
\end{eqnarray}
}

{
For the many electrons state ($\mu ,\nu = 0,1,2,3,\cdots ,N$),
the size of $\bar{K}$, $\bar{L}$, $\bar{M}$, and $\bar{N}$ is $(N+1)^2 \times (N+1)^2$.
In case of zero temperature, the consideration can be reduced due to $\rho_{0, \mu} = \delta_{\mu ,0}$;
hence, the matrix size is $(N+1) \times (N+1)$.
For the separable kernel of delta function, contrarily, if one considers $M$ bases ($i,j = 1,2,3,\cdots ,M$), $\bar{R}$ is $3M \times 3M$.
Here, the factor $3$ comes from $\alpha = x,y,z$.
In addition, $\bar{U}$ and $\bar{V}$ are $(N+1)^2 \times 3M$ (or $(N+1) \times 3M$),
and $\bar{S}$ and $\bar{T}$ are $3M \times (N+1)^2$ (or $3M \times (N+1)$).
Note that the physical dimension of the matrices is not identical.
Hence, one has to modify the equations (\ref{eq:exSCmatfrom1})-(\ref{eq:exSCmatfrom3}).
}

{
Equation (\ref{eq:exSCmatfrom3}) is deformed as
\begin{equation}
\bm{X}^{(A)} = \frac{1}{\bar{1} - \bar{R}}
\left[ \bm{Y}^{(A)} - \bar{S} \bm{X}^{(-)} - \bar{T} \bm{X}^{(+)} \right].
\end{equation}
A substitution to Eqs.\ (\ref{eq:exSCmatfrom1}) and (\ref{eq:exSCmatfrom2}) gives
\begin{eqnarray}
\left[ \left( \hbar \bar{\Omega} - (\hbar \omega + i\gamma) \bar{1} \right) + \bar{K}
+ \bar{U} \frac{1}{\bar{1} - \bar{R}} \bar{S} \right] \bm{X}^{(-)}
+ \left[ \bar{L} + \bar{U} \frac{1}{\bar{1} - \bar{R}} \bar{T} \right] \bm{X}^{(+)}
&=&
\bm{Y}^{(0,-)} + \bar{U} \frac{1}{\bar{1} - \bar{R}} \bm{Y}^{(A)},
\\
\left[ \bar{M} + \bar{V} \frac{1}{\bar{1} - \bar{R}} \bar{S} \right] \bm{X}^{(-)}
+ \left[ \left( \hbar \bar{\Omega} + (\hbar \omega + i\gamma) \bar{1} \right) + \bar{N}
+ \bar{V} \frac{1}{\bar{1} - \bar{R}} \bar{T} \right] \bm{X}^{(+)}
&=&
\bm{Y}^{(0,+)} + \bar{V} \frac{1}{\bar{1} - \bar{R}} \bm{Y}^{(A)}.
\end{eqnarray}
As a result, one obtains Eq.\ (\ref{eq:exSCmat}) with Eq.\ (\ref{eq:exSCmatXi}) and (\ref{eq:exSCmatY}):
\begin{equation}
\left[ \bar{\Xi} (\omega) \right]
\left( \begin{matrix}
\bm{X}^{(-)} \\
\bm{X}^{(+)}
\end{matrix} \right)
=
\left( \begin{matrix}
\bm{Z}^{(0,-)} \\
\bm{Z}^{(0,+)}
\end{matrix} \right)
\end{equation}
with
\begin{equation}
\bar{\Xi}
=
\left[
\left( \begin{matrix}
\hbar \bar{\Omega} - (\hbar \omega + i\gamma) \bar{1} &
 \\
 &
 \hbar \bar{\Omega} + (\hbar \omega + i\gamma) \bar{1} \\
\end{matrix} \right)
+
\left( \begin{matrix}
\bar{K} & \bar{L} \\
\bar{M} & \bar{N}
\end{matrix} \right)
+
\left( \begin{matrix}
\bar{U} \frac{1}{\bar{1} - \bar{R}} \bar{S} &
\bar{U} \frac{1}{\bar{1} - \bar{R}} \bar{T} \\
\bar{V} \frac{1}{\bar{1} - \bar{R}} \bar{S} &
\bar{V} \frac{1}{\bar{1} - \bar{R}} \bar{T}
\end{matrix} \right)
\right]
\end{equation}
and
\begin{equation}
\left( \begin{matrix}
\bm{Z}^{(0,-)} \\
\bm{Z}^{(0,+)}
\end{matrix} \right)
=
\left( \begin{matrix}
\bm{Y}^{(0,-)} \\
\bm{Y}^{(0,+)}
\end{matrix} \right)
+
\left( \begin{matrix}
\bar{U} \frac{1}{\bar{1} - \bar{R}} \bm{Y}^{(A)} \\
\bar{V} \frac{1}{\bar{1} - \bar{R}} \bm{Y}^{(A)}
\end{matrix} \right).
\end{equation}
}

{
\section{Radiative correction for rectangular nanorod}
\label{appx:NanoRod}
In this appendix, we summarize the detailed calculations for the rectangular nanorod discussed in Sec.\ \ref{sec:XimatrixNR}.
}

{
Single particle wavefunction for the electron and hole are given by Eqs.\ (\ref{eq:wavefunc_e}) and (\ref{eq:wavefunc_h}).
\begin{eqnarray}
\psi_{{\rm e}\mu} (\bm{x})
&=& \sqrt{\frac{2}{L_x}} \sin \left( \frac{n_x \pi}{L_x} x \right)
    \sqrt{\frac{2}{L_y}} \sin \left( \frac{n_y \pi}{L_y} y \right)
    \sqrt{\frac{2}{L_z}} \sin \left( \frac{n_z \pi}{L_z} z \right),
\\
\psi_{{\rm h}\bar{\mu}} (\bm{x})
&=& \sqrt{\frac{2}{L_x}} \sin \left( \frac{\bar{n}_x \pi}{L_x} x \right)
    \sqrt{\frac{2}{L_y}} \sin \left( \frac{\bar{n}_y \pi}{L_y} y \right)
    \sqrt{\frac{2}{L_z}} \sin \left( \frac{\bar{n}_z \pi}{L_z} z \right).
\end{eqnarray}
For these wavefunctions, the excited charge and current densities at zero temperature,
$\rho_{0\mu} (\bm{x})$ and $\bm{j}_{0\mu} (\bm{x})$, are obtained by Eqs.\ (\ref{eq:rho_he}) and (\ref{eq:j_he})
with $\mu = (\rm{e}\mu ,\rm{h}\bar{\mu})$:
\begin{eqnarray}
(c/n_{\rm bg}) \rho_{0\mu} (\bm{x})
&=&
\frac{(c/n_{\rm bg}) e}{L_x L_y L_z}
\left\{ \cos \Big( (q_{\bar{n}_x} - q_{n_x}) x \Big) - \cos \Big( (q_{\bar{n}_x} + q_{n_x}) x \Big) \right\}
\nonumber \\
& & \hspace{10mm} \times
\left\{ \cos \Big( (q_{\bar{n}_y} - q_{n_y}) y \Big) - \cos \Big( (q_{\bar{n}_y} + q_{n_y}) y \Big) \right\}
\nonumber \\
& & \hspace{10mm} \times
\left\{ \cos \Big( (q_{\bar{n}_z} - q_{n_z}) z \Big) - \cos \Big( (q_{\bar{n}_z} + q_{n_z}) z \Big) \right\},
\label{eq:rhohe_x} \\
j_{0\mu}^{(x)} (\bm{x})
&=&
-\frac{e\hbar}{2im_{\rm e}} \frac{1}{L_x L_y L_z} \left[
q_{\bar{n}_x} \left\{ \sin \Big( (q_{\bar{n}_x} + q_{n_x}) x \Big) - \sin \Big( (q_{\bar{n}_x} - q_{n_x}) x \Big) \right\}
\right.
\nonumber \\
& & \hspace{27mm}
\left.
- q_{n_x}     \left\{ \sin \Big( (q_{\bar{n}_x} + q_{n_x}) x \Big) + \sin \Big( (q_{\bar{n}_x} - q_{n_x}) x \Big) \right\}
\right]
\nonumber \\
& & \hspace{18mm} \times
\left\{ \cos \Big( (q_{\bar{n}_y} - q_{n_y}) y \Big) - \cos \Big( (q_{\bar{n}_y} + q_{n_y}) y \Big) \right\}
\nonumber \\
& & \hspace{18mm} \times
\left\{ \cos \Big( (q_{\bar{n}_z} - q_{n_z}) z \Big) - \cos \Big( (q_{\bar{n}_z} + q_{n_z}) z \Big) \right\},
\label{eq:jhe_x}
\end{eqnarray}
and similar manner for $j_{0\mu}^{(y)}$ and $j_{0\mu}^{(z)}$.
Here, $q_{n_\alpha} = \pi n_\alpha /L_\alpha$ for $\alpha = x,y,z$.
Note that an assumed background refractive index $n_{\rm bg}$ modulates the light velocity to $c/n_{\rm bg}$.
For the calculation, we use the $\bm{k}$-representation. Then,
\begin{eqnarray}
(c/n_{\rm bg}) \tilde{\rho}_{0\mu} (\bm{k})
&=&
\frac{1}{(2\pi)^{3/2}} \int d\bm{x} (c/n_{\rm bg}) \rho_{0\mu} (\bm{x}) e^{-i\bm{k} \cdot \bm{x}}
\nonumber \\
&=&
\frac{(c/n_{\rm bg}) e}{8L_xL_yL_z} \frac{1}{(2\pi)^{3/2}} \frac{1}{i^3}
\nonumber \\
& & \hspace{3mm} \times
\left[ \frac{2k_x}{{k_x}^2 - {\Delta_{n_x,\bar{n}_x}}^2} - \frac{2k_x}{{k_x}^2 - {Q_{n_x,\bar{n}_x}}^2} \right]
\left[ \frac{2k_y}{{k_y}^2 - {\Delta_{n_y,\bar{n}_y}}^2} - \frac{2k_y}{{k_y}^2 - {Q_{n_y,\bar{n}_y}}^2} \right]
\left[ \frac{2k_z}{{k_z}^2 - {\Delta_{n_z,\bar{n}_z}}^2} - \frac{2k_z}{{k_z}^2 - {Q_{n_z,\bar{n}_z}}^2} \right]
\nonumber \\
& & \hspace{3mm} \times
\left( 1 - (-1)^{N_x} e^{-ik_xL_x} \right)
\left( 1 - (-1)^{N_y} e^{-ik_yL_y} \right)
\left( 1 - (-1)^{N_z} e^{-ik_zL_z} \right)
\label{eq:rhohe_k},
\\
\tilde{j}_{0\mu}^{(x)} (\bm{k})
&=&
\frac{1}{(2\pi)^{3/2}} \int d\bm{x} j_{0\mu}^{(x)} (\bm{x}) e^{-i\bm{k} \cdot \bm{x}}
\nonumber \\
&=&
\frac{e\hbar}{2m_{\rm e}} \frac{1}{8L_xL_yL_z} \frac{1}{(2\pi)^{3/2}} \frac{1}{i^3}
\nonumber \\
& & \hspace{3mm} \times
\left[ \frac{2\Delta_{n_x,\bar{n}_x} Q_{n_x,\bar{n}_x}}{{k_x}^2 - {\Delta_{n_x,\bar{n}_x}}^2}
     - \frac{2Q_{n_x,\bar{n}_x} \Delta_{n_x,\bar{n}_x}}{{k_x}^2 - {     Q_{n_x,\bar{n}_x}}^2} \right]
\left[ \frac{2k_y}{{k_y}^2 - {\Delta_{n_y,\bar{n}_y}}^2} - \frac{2k_y}{{k_y}^2 - {Q_{n_y,\bar{n}_y}}^2} \right]
\left[ \frac{2k_z}{{k_z}^2 - {\Delta_{n_z,\bar{n}_z}}^2} - \frac{2k_z}{{k_z}^2 - {Q_{n_z,\bar{n}_z}}^2} \right]
\nonumber \\
& & \hspace{3mm} \times
\left( 1 - (-1)^{N_x} e^{-ik_xL_x} \right)
\left( 1 - (-1)^{N_y} e^{-ik_yL_y} \right)
\left( 1 - (-1)^{N_z} e^{-ik_zL_z} \right)
\label{eq:jhe_k}
\end{eqnarray}
with following definitions: $N_\alpha \equiv \bar{n}_\alpha + n_\alpha$,
$Q_{n_\alpha,\bar{n}_\alpha} \equiv q_{n_\alpha} + q_{\bar{n}_\alpha}$, and
$\Delta_{n_\alpha,\bar{n}_\alpha} \equiv q_{n_\alpha} - q_{\bar{n}_\alpha}$.
Moreover, we find symmetric relations,
$\tilde{\rho}_{\mu0} (\bm{k})   =   \tilde{\rho}_{0\mu} (\bm{k})$ and
$\tilde{\bm{j}}_{\mu0} (\bm{k}) = - \tilde{\bm{j}}_{0\mu} (\bm{k})$.
}

{
The elements of the radiative correction matrix $\bar{K}$ in Eq.\ (\ref{eq:Kmu00mu}) is
\begin{equation}
\tilde{K}_{\mu' 0,0\mu} = A_{\mu',\mu}^{(2)} (\omega) + A_{\mu',\mu}^{(1)} (\omega) + A_{\mu',\mu}^{(0)}
\end{equation}
with
\begin{eqnarray}
A_{\mu',\mu}^{(0)}
&=&
\mu_0 \int d\bm{k}
(c/n_{\rm bg}) \tilde{\rho}_{\mu'0}(-\bm{k}) \left( \frac{1}{\bm{k}^2} \right) (c/n_{\rm bg}) \tilde{\rho}_{0\mu}(\bm{k})
\nonumber \\
&=&
\mu_0 \left( \frac{{L_0}^3}{L_x L_y L_z} \right)^2 \left(\frac{2}{\pi} \right)^3 \frac{c^2 e^2}{\pi^5 L_0}
\cdot
\int_{-\infty}^{\infty}dX \int_{-\infty}^{\infty}dY \int_{-\infty}^{\infty}dZ
\nonumber \\
& & \hspace{10mm} \times
\left( X^2 \bar{F}_{n'_x,\bar{n'}_x;n_x,\bar{n}_x} (X) \right)
\left( Y^2 \bar{F}_{n'_y,\bar{n'}_y;n_y,\bar{n}_y} (Y) \right)
\left( Z^2 \bar{F}_{n'_z,\bar{n'}_z;n_z,\bar{n}_z} (Z) \right)
\frac{1}{X^2 + Y^2 + Z^2} \frac{1}{{n_{\rm bg}}^2}
\label{eq:A0_nodim_X} \\
A_{\mu',\mu}^{(1)} (\omega)
&=&
\mu_0 \int d\bm{k} \left(\tilde{\bm{j}}_{\mu' 0}(-\bm{k}) \right)^{\rm t}
\left( - \frac{1}{\bm{k}^2} \frac{n_{\rm bg} \omega/c}{- \bm{k}^2 + (n_{\rm bg} \omega/c)^2} \bm{k} \right)
(c/n_{\rm bg}) \tilde{\rho}_{0 \mu}(\bm{k})
\nonumber \\
&=&
\eta_\omega \mu_0 \left( \frac{{L_0}^3}{L_x L_y L_z} \right)^2 \left(\frac{2}{\pi} \right)^3 \frac{c^2 e^2}{\pi^5 L_0}
\cdot
\int_{-\infty}^{\infty}dX \int_{-\infty}^{\infty}dY \int_{-\infty}^{\infty}dZ
\nonumber \\
& & \hspace{10mm} \times
\left( \bar{Q}_{n'_x,\bar{n'}_x}\bar{\Delta}_{n'_x,\bar{n'}_x}
     + \bar{Q}_{n'_y,\bar{n'}_y}\bar{\Delta}_{n'_y,\bar{n'}_y}
     + \bar{Q}_{n'_z,\bar{n'}_z}\bar{\Delta}_{n'_z,\bar{n'}_z} \right)
\nonumber \\
& & \hspace{10mm} \times
\left( X^2 \bar{F}_{n'_x,\bar{n'}_x;n_x,\bar{n}_x} (X) \right)
\left( Y^2 \bar{F}_{n'_y,\bar{n'}_y;n_y,\bar{n}_y} (Y) \right)
\left( Z^2 \bar{F}_{n'_z,\bar{n'}_z;n_z,\bar{n}_z} (Z) \right)
\nonumber \\
& & \hspace{60mm} \times
\frac{1}{X^2 + Y^2 + Z^2} \frac{1}{(X^2 + Y^2 + Z^2) - {n_{\rm bg}}^2 \bar{\omega}^2}
\label{eq:A1_nodim_X} \\
A_{\mu',\mu}^{(2)} (\omega)
&=&
\mu_0 \int d\bm{k} \left(\tilde{\bm{j}}_{\mu' 0}(-\bm{k}) \right)^{\rm t}
\left( \frac{1}{- \bm{k}^2 + (n_{\rm bg} \omega/c)^2} \right) \tilde{\bm{j}}_{0\mu}(\bm{k})
\nonumber \\
&=&
- \frac{{\eta_\omega}^2}{\bar{\omega}^2} \mu_0 \left( \frac{{L_0}^3}{L_x L_y L_z} \right)^2
\left(\frac{2}{\pi} \right)^3 \frac{c^2 e^2}{\pi^5 L_0} \cdot
\int_{-\infty}^{\infty}dX \int_{-\infty}^{\infty}dY \int_{-\infty}^{\infty}dZ
\nonumber \\
& & \hspace{10mm} \times
\left[
\left( \bar{\Pi}_{n'_x,\bar{n'}_x;n_x,\bar{n}_x} \bar{F}_{n'_x,\bar{n'}_x;n_x,\bar{n}_x} (X) \right)
\left( Y^2                                       \bar{F}_{n'_y,\bar{n'}_y;n_y,\bar{n}_y} (Y) \right)
\left( Z^2                                       \bar{F}_{n'_z,\bar{n'}_z;n_z,\bar{n}_z} (Z) \right)
\right.
\nonumber \\
& & \hspace{10mm}
+
\left( X^2                                       \bar{F}_{n'_x,\bar{n'}_x;n_x,\bar{n}_x} (X) \right)
\left( \bar{\Pi}_{n'_y,\bar{n'}_y;n_y,\bar{n}_y} \bar{F}_{n'_y,\bar{n'}_y;n_y,\bar{n}_y} (Y) \right)
\left( Z^2                                       \bar{F}_{n'_z,\bar{n'}_z;n_z,\bar{n}_z} (Z) \right)
\nonumber \\
& & \hspace{10mm}
\left. +
\left( X^2                                       \bar{F}_{n'_x,\bar{n'}_x;n_x,\bar{n}_x} (X) \right)
\left( Y^2                                       \bar{F}_{n'_y,\bar{n'}_y;n_y,\bar{n}_y} (Y) \right)
\left( \bar{\Pi}_{n'_z,\bar{n'}_z;n_z,\bar{n}_z} \bar{F}_{n'_z,\bar{n'}_z;n_z,\bar{n}_z} (Z) \right)
\right]
\nonumber \\
& & \hspace{70mm} \times
\frac{1}{(X^2 + Y^2 + Z^2) - {n_{\rm bg}}^2 \bar{\omega}^2}.
\label{eq:A2_nodim_X}
\end{eqnarray}
Here, the functions for the integrands are
\begin{eqnarray}
\bar{F}_{n'_x,\bar{n'}_x;n_x,\bar{n}_x} (X)
&=&
\left( \frac{1}{X^2 - {\bar{\Delta}_{n'_x,\bar{n'}_x}}^2} - \frac{1}{X^2 - {\bar{Q}_{n'_x,\bar{n'}_x}}^2} \right)
\left( \frac{1}{X^2 - {\bar{\Delta}_{n_x ,\bar{n}_x }}^2} - \frac{1}{X^2 - {\bar{Q}_{n_x ,\bar{n}_x }}^2} \right)
\nonumber \\
& & \hspace{10mm} \times
\frac{1}{4} \left( 1 + (-1)^{N_x + N'_x} - (-1)^{N_x} e^{-i\pi L_x X/L_0} - (-1)^{N'_x} e^{i\pi L_x X/L_0} \right)
\label{eq:XF_nodim_X} \\
\bar{\Pi}_{n'_\alpha,\bar{n'}_\alpha;n_\alpha,\bar{n}_\alpha}
&=&
\bar{Q}_{n'_\alpha,\bar{n'}_\alpha} \bar{\Delta}_{n'_\alpha,\bar{n'}_\alpha}
\bar{Q}_{n_\alpha ,\bar{n}_\alpha } \bar{\Delta}_{n_\alpha ,\bar{n}_\alpha }.
\label{eq:PiQDQD_X}
\end{eqnarray}
$\bar{F}_{n'_y,\bar{n'}_y;n_y,\bar{n}_y} (Y)$ and $\bar{F}_{n'_z,\bar{n'}_z;n_z,\bar{n}_z} (Z)$ are defined in
a similar manner for Eq.\ (\ref{eq:XF_nodim_X}).
The integral variables and several factors are normalized as dimensionless ones,
$X = k_x L_0/\pi$, $Y = k_y L_0/\pi$, $Z = k_z L_0/\pi$,
$\bar{Q}_{n_\alpha ,\bar{n}_\alpha} = Q_{n_\alpha ,\bar{n}_\alpha} L_0/\pi$,
$\bar{\Delta}_{n_\alpha ,\bar{n}_\alpha} = \Delta_{n_\alpha ,\bar{n}_\alpha} L_0/\pi$, and
\begin{equation}
\bar{\omega} = \frac{\omega}{c} \frac{L_0}{\pi}.
\label{eq:dess_omega}
\end{equation}
Here, an introduced factor
\begin{equation}
\eta_\omega = \frac{\pi}{2} \frac{\hbar}{m^* c L_0} \bar{\omega}
= \frac{1}{m^* /m_{\rm e}} \frac{\hbar \omega}{2m_{\rm e} c^2}
\label{eq:relativistic}
\end{equation}
means a relativistic factor.
The elements of the matrices $\bar{L}$, $\bar{M}$, and $\bar{N}$ are obtained by Eqs.\ (\ref{eq:Lmu0mu0}),
(\ref{eq:M0mu0mu}), and (\ref{eq:N0mumu0}), respectively.
}

{
For the matrices related to the vector potential term in the current,
the elements of $\bar{S}$ in Eq.\ (\ref{eq:Smbmu0}) becomes
\begin{equation}
\tilde{S}_{m'\beta,0 \mu} = B_{m'\beta,\mu}^{(2)} + B_{m'\beta,\mu}^{(1)}
\end{equation}
with
\begin{eqnarray}
B_{m'x,\mu}^{(1)} (\omega)
&=&
\mu_0 \int d\bm{k} \tilde{\varphi}_{m'} (-\bm{k})
\left(-\frac{1}{\bm{k}^2} \frac{n_{\rm bg} \omega/c}{- \bm{k}^2 + (n_{\rm bg} \omega/c)^2} k_x \right)
(c/n_{\rm bg}) \tilde{\rho}_{0\mu} (\bm{k})
\nonumber \\
&=&
-i\mu_0 \left(\frac{{L_0}^3}{L_x L_y L_z} \right)^{\frac{3}{2}} \left( \frac{2}{\pi} \right)^{\frac{9}{2}}
\left( \frac{c^2 e^2}{\pi^5 L_0} \right)^{\frac{1}{2}} \frac{\omega {L_0}^2}{c \pi^2}
\nonumber \\
& & \hspace{10mm} \times
\int_{-\infty}^{\infty}dX \int_{-\infty}^{\infty}dY \int_{-\infty}^{\infty}dZ
\left( X^2 \bar{h}_{m'_x;n_x,\bar{n}_x} (X) \right)
\left(   Y \bar{h}_{m'_y;n_y,\bar{n}_y} (Y) \right)
\left(   Z \bar{h}_{m'_z;n_z,\bar{n}_z} (Z) \right)
\nonumber \\
& & \hspace{60mm} \times
\frac{1}{X^2 + Y^2 + Z^2}\frac{1}{(X^2 + Y^2 + Z^2) - {n_{\rm bg}}^2 \bar{\omega}^2},
\label{eq:B1_nodim_X} \\
B_{m'x,\mu}^{(2)} (\omega)
&=&
\mu_0 \int d\bm{k} \tilde{\varphi}_{m'} (-\bm{k})
\left(\frac{1}{- \bm{k}^2 + (n_{\rm bg} \omega/c)^2} \right) \tilde{j}^{(x)}_{0\mu} (\bm{k})
\nonumber \\
&=&
i \left( \frac{\eta_\omega}{\bar{\omega}^2} \right)
\mu_0 \left( \frac{{L_0}^3}{L_x L_y L_z} \right)^{\frac{3}{2}} \left( \frac{2}{\pi} \right)^{\frac{9}{2}}
\left( \frac{c^2 e^2}{\pi^5 L_0} \right)^{\frac{1}{2}} \frac{\omega {L_0}^2}{c \pi^2}
\nonumber \\
& & \hspace{10mm} \times
\int_{-\infty}^{\infty}dX \int_{-\infty}^{\infty}dY \int_{-\infty}^{\infty}dZ
\left( \bar{\pi}_{n_x,\bar{n}_x} \bar{h}_{m'_x;n_x,\bar{n}_x} (X) \right)
\left(                         Y \bar{h}_{m'_y;n_y,\bar{n}_y} (Y) \right)
\left(                         Z \bar{h}_{m'_z;n_z,\bar{n}_z} (Z) \right)
\nonumber \\
& & \hspace{60mm} \times
\frac{1}{(X^2 + Y^2 + Z^2) - {n_{\rm bg}}^2 \bar{\omega}^2}.
\label{eq:B2_nodim_X}
\end{eqnarray}
Here, the functions for the integrands are
\begin{eqnarray}
X \bar{h}_{m'_x;n_x,\bar{n}_x} (X)
&=&
\frac{\bar{Q}_{m'_x} }{X^2 - {\bar{Q}_{m'_x}^2} }
\left[ \frac{X}{X^2 - {\bar{\Delta}_{n_x,\bar{n}_x}}^2} - \frac{X}{X^2 - {\bar{Q}_{n_x,\bar{n}_x}}^2} \right]
\nonumber \\
& & \hspace{15mm} \times
\frac{1}{4} \left( 1 + (-1)^{N_x + m'_x} - (-1)^{N_x} e^{-i\pi L_x X/L_0} - (-1)^{m'_x} e^{i\pi L_x X/L_0} \right),
\label{eq:Xh_X} \\
\bar{\pi}_{n_\alpha,\bar{n}_\alpha}
&=&
\bar{Q}_{n_\alpha \bar{n}_\alpha} \bar{\Delta}_{n_\alpha,\bar{n}_\alpha},
\label{eq:piQD_nodim_X}
\end{eqnarray}
with a dimensionless factor $\bar{Q}_{m_\alpha} = q_{m_\alpha} L_0 /\pi$.
$B_{m'y,\mu}^{(j)} (\omega)$ and $B_{m'z,\mu}^{(j)} (\omega)$ are obtained in the similar manner.
The elements of $\bar{T}$, $\bar{U}$, and $\bar{V}$ in Eqs.\ (\ref{eq:Tmb0mu})-(\ref{eq:V0muma}) are also given by them.
}

{
Finally, for the matrix $\bar{R}$, Eq.\ (\ref{eq:Rmbma}) becomes
\begin{eqnarray}
C_{m',m} (\omega)
&=&
\left( \frac{\omega_{\rm p}^\prime}{c/n_{\rm bg}} \right)^2 \int d^3\bm{k} \tilde{\varphi}_{m'} (-\bm{k})
\left( \frac{1}{- \bm{k}^2 + (n_{\rm bg} \omega/c)^2} \right) \tilde{\varphi}_m^\ast (\bm{k}) \delta_{\alpha \beta}
\nonumber \\
&=&
- \frac{{L_0}^3}{L_x L_y L_z} \left(\frac{2}{\pi}\right)^6 \left( \frac{\omega_{\rm p}}{\omega} \right)^2
\left( \frac{\omega L_0}{c \pi} \right)^2
\cdot
\int_{-\infty}^{\infty}dX \int_{-\infty}^{\infty}dY \int_{-\infty}^{\infty}dZ
\nonumber \\
& & \hspace{10mm} \times
\left(\bar{f}_{m'_x ;m_x} (X) \right)
\left(\bar{f}_{m'_y ;m_y} (Y) \right)
\left(\bar{f}_{m'_z ;m_z} (Z) \right)
\frac{1}{(X^2 + Y^2 + Z^2) - {n_{\rm bg}}^2 \bar{\omega}^2}
\label{eq:C_nodim_X}
\end{eqnarray}
with
\begin{equation}
\bar{f}_{m'_x;m_x} (X) =
\frac{\bar{Q}_{m'_x}}{X^2 - {\bar{Q}_{m'_x}^2}} \frac{\bar{Q}_{m_x} }{X^2 - {\bar{Q}_{m_x}^2} }
\times \frac{1}{4} \left( 1 + (-1)^{m_x + m'_x} - (-1)^{m_x} e^{-i\pi L_x X/L_0} - (-1)^{m'_x} e^{i\pi L_x X/L_0} \right).
\label{eq:f_nodim_X}
\end{equation}
}


\begin{thebibliography}{99}
\bibitem{com1}
Complete inclusion of the correlation and quantum fluctuation for the excited states
is a task beyond the simple preparation of the exact eigenstates $|\mu \rangle$ including the 
Coulomb interaction at equilibrium. This point will be a subject of future research based on the present study.
\end{thebibliography}
\end{document}